\documentclass[journal]{IEEEtran}

\usepackage{epsf}
\usepackage{graphicx}
\DeclareGraphicsExtensions{.png}
\usepackage{caption}
\usepackage{subcaption}
\usepackage{xcolor}
\usepackage{amsmath} \interdisplaylinepenalty=2500
\usepackage{amsfonts}
\usepackage{amssymb}
\usepackage{array}
\usepackage{cuted}

\newcommand{\ul}[1]{\underline{#1}}

\newcommand{\eps}{\epsilon}

\newcommand{\bea}{\begin{eqnarray}}

\newcommand{\eea}{\end{eqnarray}}

\newcommand{\rmc}{{\rm c}}

\newcommand{\rmd}{{\rm d}}

\newcommand{\rme}{{\rm e}}

\newcommand{\rmem}{{\rm em}}

\newcommand{\rmH}{{\rm H}}

\newcommand{\rmj}{{\rm j}}

\newcommand{\rmm}{{\rm m}}

\newcommand{\cE}{{\cal E}}

\newcommand{\cF}{{\cal F}}

\newcommand{\cG}{{\cal G}}

\newcommand{\cH}{{\cal H}}

\newcommand{\Ups}{{\mit\Upsilon}}

\begin{document}



\title{Fluctuations of Power versus Energy for Random Fields Near a Perfectly Conducting Boundary
}

\author{
{Luk~R.~Arnaut
}%
}



\maketitle



\begin{abstract}
The standard deviations of the energy and Poynting power densities for an isotropic random field near a perfectly conducting planar boundary are characterized, based on quartic plane-wave expansions. For normal and transverse components, different rates of decay exist as a function of electrical distance from the boundary. At large distances, the envelopes for the power are more strongly damped than for the energy, both showing inverse power law decay. The decay for the standard deviation is generally one order faster than for the corresponding mean. For the normally directed power flux, its standard deviation near the boundary increases linearly with distance. The relative uncertainty of the scalar power is much smaller than for the Poynting power. Poynting's theorem for standard deviations is obtained and demonstrates larger standard deviations of the energy imbalance and power flux than their mean values. 
\end{abstract}


\section{Introduction\label{sec:intro}}
\IEEEPARstart{T}{he} 
 characterization of the fluctuations of complex random fields, energy, and power near an electromagnetic (EM) boundary is an aspect of fundamental importance.
Some recent studies include the evaluation of linear and orbital angular momentum near a ground plane \cite{arnaLMAM}, the probing of fields near a metal wall in a mode-stirred reverberation chamber (MSRC) \cite{deleo2020}, and the propagation of uncertainty and correlation in computational EM field solvers \cite{smith2012}. 

When measuring deterministic fields, practitioners have a choice between electric (or magnetic) field sensors of wire dipole (or loop) type versus aperture antennas or waveguides that measure EM power flow. The choice is largely governed by the sensitivity at the frequency of interest, the size and practicality. For plane waves, the conversion from voltage, field intensity, electric (or magnetic) energy to EM power is straightforward. 
 
Random fields near an EM boundary prompt a consideration of the effects of the boundary conditions on the root-mean-square (RMS) fluctuations (standard deviation) of power vs. energy. Any such effects complement those that have been found to exist for their expected values (averages) \cite{arnaLMAM}, which governs the optimal sensor placement for eliminating the effect of the boundary. 
As will be shown here, the choice of sensor type and measurand leads to supplementary differences in the signal-to-noise ratio (SNR) between random energy vs. random power that affect their relative uncertainties. 

In \cite{hill1998}, the issue of defining power in an ideal reverberant space was raised.
While the {\em average\/} local power is null across all angles of incidence ($4\pi$ sr), this does not preclude significant nonzero fluctuations of power that may still produce physical effects and require characterization. 
Near a perfect electric conducting (PEC) boundary, the average power \cite{arnaLMAM} and energy \cite{dunn1990}--\cite{arnaRS2007} of random fields are both nonzero and exhibit different spatial dependencies, through decaying interference at-a-distance, unlike for deterministic fields. 
One consequence is the recommendation to avoid using a boundary layer zone of quarter-wavelength thickness near any metallic surface inside a MSRC \cite{iec}. However, this guard distance creates practical problems in testing heavy floor-standing equipment or inside small cabinets. More generally, it is wasteful of the available interior space, particularly at relatively low frequencies. 
Some studies have started to address this issue \cite[sec. III-C]{arnaLMAM}, \cite{arnaTEMCmay2006}, \cite[sec. 5.3.5]{arnaTQE2}, \cite{serr2020}, with a view to increase the working volume of a MSRC up to its physical boundaries. To this end, the full statistical characterization of the EM energy and power in the boundary zone is required. 

In this article, the work in \cite{arnaLMAM}, \cite{dunn1990}, \cite{hill2005}, \cite{arnaTEMCmay2006} on average energy and Poynting power densities for random fields near a PEC boundary is extended to cover corresponding standard deviations. A time-harmonic dependence $\exp(\rmj \omega t)$ is assumed and suppressed. To simplify notation, the argument of all spherical Bessel functions will be dropped, i.e., $j_\ell \equiv j_\ell(2kz)$ for $\ell=0,1,2$, which are hence always evaluated at $2kz$. 

\section{Spectral Plane Wave Expansions for Quartic Mixed Products of Circular Gaussian Fields}
The angular spectral plane-wave expansion is defined by
\begin{align}
\ul{F} (\ul{r}) &= \frac{1}{\Omega} \iint_\Omega \ul{\cF}(\Omega) \exp (-\rmj \ul{k} \cdot \ul{r}) \rmd\Omega
\label{eq:Ealpha}
\end{align}
where $\ul{F}=\ul{E}$ or $\ul{H}$ and $\ul{{\cal F}}=\ul{{\cal E}}$ or $\ul{{\cal H}}$, which can be traced to Cram\'{e}r's spectral representation of scalar stationary random functions \cite{dunn1990}, \cite{yagl1962}.
For the time-averaged local electric and magnetic energy densities, $U_{\rme}(\ul{r}) = \eps_0 \ul{E}(\ul{r}) \cdot \ul{E}^*(\ul{r}) / 4$ and $U_{\rmm}(\ul{r}) = \mu_0 \ul{H}(\ul{r}) \cdot \ul{H}^*(\ul{r}) / 4$, and for the Poynting power vector $\ul{S}(\ul{r}) = [\ul{E}(\ul{r}) \times \ul{H}^*(\ul{r})]/2$, the expansions involve two-factor pure or mixed products of plane wave fields, viz., \cite{arnaLMAM}
\begin{align}
F_\alpha (\ul{r}) G^*_\beta (\ul{r}) 
&=
\frac{1}{\Omega}
\left ( \iint_\Omega \right )^2 {\cF}_{1,\alpha}
{\cG}^*_{2,\beta} 
\,\delta(\Omega_{12}) 
\nonumber\\
&~ \times
\exp \left [ -\rmj ( \ul{k}_1 - \ul{k}^*_2 ) \cdot \ul{r} \right ] 
\rmd\Omega_1 \rmd\Omega_2
\label{eq:EalphaHbeta_bis}
\end{align}
for ${F},{G} \in \{ {E},{H} \}$; ${\cF},{\cG} \in \{ {\cE},{\cH} \}$; $\alpha,\beta \in \{x,y,z\}$, and where $\cF_{i,\alpha} \stackrel{\Delta}{=} \ul{\cF}(\Omega_i) \cdot \ul{1}_\alpha$, etc., are Cartesian field components along the direction of $\ul{1}_\alpha$ for the angular spectral plane-wave components $(\ul{\cE}(\Omega_i), \ul{\cH}(\Omega_i),\ul{k}_i)$ of the observable random EM field $(\ul{E}, \ul{H})$ at location of incidence $\ul{r}$.
The wave vector $\ul{k}_i \stackrel{\Delta}{=} \ul{k}(\Omega_i)$ is quasi-monochromatic ($|\ul{k}_i| = \omega_i \sqrt{\mu_0 \eps_0}$). The analysis assumes a lossless medium of incidence ($\cE_i / \cH_i = \cE_0 / \cH_0 = \eta_0 = \sqrt{\mu_0/\eps_0}$; $\ul{k}_i = \ul{k}^*_i$).
The symmetric set difference $\Omega_{ij}\stackrel{\Delta}{=} \Omega_i \Delta \Omega_j = (\Omega_i\cup\Omega_j)\setminus(\Omega_i\cap\Omega_j)$ 
for overlap between solid angles $\Omega_{i}$ and $\Omega_{j}$ satisfies 
 $\delta (\Omega_{ij}=\emptyset)=1$ and $\delta (\Omega_{ij}\not=\emptyset)=0$.
The differentials $\rmd \Omega_i=\sin\theta_i \rmd \theta_i \rmd \phi_i$ refer to standard spherical angles of azimuth $\phi_i$ and elevation $\theta_i$, measured from zenith, for integration across the upper hemisphere $\Omega = 2\pi$ sr ($z\geq 0$).

The second-order expansions (\ref{eq:EalphaHbeta_bis}) suffice for the calculation of ensemble averages (expected values) of energy and power, denoted by $\langle \cdot \rangle$. Using a TE/TM decomposition and enforcing the EM boundary conditions at the PEC plane $oxy$ \cite{arnaLMAM}, \cite{dunn1990}
\begin{align}
\langle {S}_\alpha (kz) \rangle &= \langle {S}_t (kz) \rangle 
= 0,~~\alpha=x,y  
\label{eq:avgSx}\\
\langle {S}_z(kz) \rangle &= \langle {S}(kz) \rangle
= - \rmj \frac{ 
\langle |{\cal E}_0|^2 \rangle}{\eta_0} j_1 
\label{eq:avgSz}\\
\langle U_\alpha (kz) \rangle 
&= \frac{\varphi_0 \langle |\cF_0|^2 \rangle}{3}\left ( 1 \mp j_0 \pm \frac{j_2}{2} \right ),~~\alpha=x,y  
\label{eq:avgUexyUmxy}\\
\langle U_{z} (kz) \rangle 
&= \frac{\varphi_0 \langle |\cF_0|^2 \rangle}{3}
\left ( 1 \pm j_0 \pm j_2 \right ) \label{eq:avgUezUmz}\\
\langle U_{t} (kz) \rangle 
&= \frac{2\, \varphi_0 \langle |\cF_0|^2 \rangle}{3} \left ( 1 \mp j_0 \pm \frac{j_2}{2} \right ) 
\label{eq:avgUetUmt}\\
\langle U (kz) \rangle 
&= \varphi_0 \langle |\cF_0|^2 \rangle \left ( 1 \mp \frac{j_0 }{3} \pm \frac{2 j_2}{3} \right )  
\label{eq:avgUeUm}
\end{align}
where upper and lower signs correspond to electric $(U=U_{\rme})$ and magnetic $(U=U_{\rmm})$ energy densities, respectively, and $\varphi_0 \langle |\cF_0|^2 \rangle \stackrel{\Delta}{=} \eps_0 \langle |\cE_0|^2 \rangle = \mu_0 \langle |\cH_0|^2 \rangle$.
The subscripts $\alpha$, $z$, and $t$ represent 1-D tangential, 1-D normal, and 2-D tangential ($xy$-) components, respectively. Absence of a subscript refers to the 3-D full (vector) field $\ul{E}$ or $\ul{H}$. The total (combined, summed) electric and magnetic energy will be denoted as $U_{\rmem}$.

The expectations (\ref{eq:avgSx})--(\ref{eq:avgUeUm}) presume a set of orthogonality conditions to hold for the second-order moments of the local plane-wave complex fields $\ul{\cF}_{i} \equiv \ul{\cF}^{\prime}_{i} -\rmj \ul{\cF}^{\prime\prime}_{i} = {\cF}_{i,\phi} \ul{1}_{\phi_i} + {\cF}_{i,\theta} \ul{1}_{\theta_i}$ in the transverse plane $o\phi_i\theta_i$, viz., \cite{hill1998}, \cite{yagl1962}
\begin{align}
C_{\cF^{\prime(\prime)}_{i,\phi} \cF^{\prime(\prime)}_{j,\phi}} 
=
C_{\cF^{\prime(\prime)}_{i,\theta} \cF^{\prime(\prime)}_{j,\theta}} = (\langle |{\cE_0|^2\rangle}/{4}) \, \delta(\Omega_{ij})
\label{eq:IIQQcov}
\end{align}
for the in-phase/in-phase (I/I) and quadrature/quadrature (Q/Q) covariances $C_{\cF^{\prime(\prime)}_{i,\cdot} \cF^{\prime(\prime)}_{j,\cdot}} = \langle \cF^{\prime(\prime)}_{i,\cdot} \cF^{\prime(\prime)}_{j,\cdot} \rangle$.
The delta correlation in (\ref{eq:IIQQcov}) is a consequence of the assumed stationarity (i.e., statistical homogeneity and isotropy) of incident $\ul{F}$ with respect to $\ul{r}$. For
the
I/Q and cross-component
covariances, 
the conditions are
\begin{align}
C_{\cF^{\prime}_{i,\phi} \cF^{\prime\prime}_{j,\phi}} 
=
C_{\cF^{\prime}_{i,\theta} \cF^{\prime\prime}_{j,\theta}} 
=
C_{\cF^{\prime(\prime)}_{i,\phi} \cF^{\prime(\prime)}_{j,\theta}} 
=
0.
\label{eq:IQcov}
\end{align}

The expansion (\ref{eq:EalphaHbeta_bis}) can be extended to quartic forms, as required
in the calculation of second-order moments of $U_{\rme}$, $U_{\rmm}$, $U_{\rmem}$ and $S$ involving pure or pairwise
mixed fourth-order moments of the fields, in the general form
\begin{align}
&\/
F_\alpha(\ul{r}) F^{*}_\beta(\ul{r}) 
G_\gamma(\ul{r}) G^{*}_\delta(\ul{r})
= \frac{1}{\Omega} \left ( \iint_\Omega \right )^4 
\cF_{1,\alpha} 
\cF^{*}_{2,\beta} 
\cG_{3,\gamma} 
\cG^{*}_{4,\delta}
\nonumber\\
&\/ \times \exp \left [ -\rmj ( \ul{k}_1 - \ul{k}^{*}_2 + \ul{k}_3 - \ul{k}^{*}_4 )\cdot\ul{r} \right ] \delta(\Omega_{1234}) \rmd\Omega_1 \rmd\Omega_2 \rmd\Omega_3 \rmd\Omega_4
\nonumber\\
\label{eq:FalphaFbetaGgammaGdelta}
\end{align}
for any selection of pairwise conjugate orthogonal field components along $\alpha, \beta, \gamma, \delta \in \{ x,y,z \}$,
with
$\delta(\Omega_{1234}) \equiv \delta(\Omega_1 \Delta \Omega_2 \Delta \Omega_3 \Delta \Omega_4) = \delta(\Omega_{12}) \delta(\Omega_{34})$.
Now, if all $\ul{\cF}_{i}$ and $\ul{\cG}_{j}$ are taken to be {\em centered circular Gaussian} fields, then $\langle \cF_{1,\alpha} \cF^{*}_{2,\beta} \cG_{3,\gamma} \cG^{*}_{4,\delta} \rangle$ can be expressed in terms of second-order moments only, without imposing additional fourth-order moment conditions, by applying Isserlis's theorem for Gaussian moments to the individual real and imaginary parts of $\cF_{i,\alpha}$ and $\cG_{j,\gamma}$. 
Expressing the kernel of (\ref{eq:FalphaFbetaGgammaGdelta}) as a sum of products of pairwise covariances for real variates, followed by a recombination to the sought covariances of the complex fields, Isserlis's theorem for the complex plane-wave field components follows as
\begin{align}
&\/ \langle \cF_{1,\alpha} \cF^{*}_{2,\beta} \cG_{3,\gamma} \cG^{*}_{4,\delta} \rangle 
 = 
\langle \cF_{1,\alpha} \cF^{*}_{2,\beta} \rangle 
\langle \cG_{3,\gamma} \cG^{*}_{4,\delta} \rangle 
\nonumber\\
&\/ 
+
\langle \cF_{1,\alpha} \cG_{3,\gamma} \rangle 
\langle \cF^{*}_{2,\beta} \cG^{*}_{4,\delta} \rangle 
+
\langle \cF_{1,\alpha} \cG^{*}_{4,\delta} \rangle 
\langle \cF^{*}_{2,\beta} \cG_{3,\gamma} \rangle 
.
\label{eq:Isserlis}
\end{align}
For the mixed fourth moment $\langle |\cF_{i,\alpha}|^2 |\cG_{j,\gamma}|^2 \rangle$, its terms in (\ref{eq:Isserlis}) contain $\langle \cF_{i,\alpha} \cG_{j,\gamma} \rangle$ and $\langle \cF_{i,\alpha} \cG^*_{j,\gamma} \rangle$, which may vanish or not \cite{arnaLMAM}, as will become apparent in the further analysis.

In summary, the conditions (\ref{eq:IIQQcov}) and (\ref{eq:IQcov}) on the second-order moments are necessary and sufficient to fully characterize all even-order moments of circular complex Gaussian $\cF_{i,\alpha}$ and $\cG_{j,\gamma}$ and hence all (even and odd) moments of energy and power, following (\ref{eq:Isserlis}). If some fields are elliptic and/or non-Gaussian, then additional conditions on their pseudovariances and/or fourth-order moments apply, respectively.

Note that Isserlis's theorem can equally be applied to $\langle F_{1,\alpha} F^{*}_{2,\beta} G_{3,\gamma} G^{*}_{4,\delta} \rangle$, i.e., just as well to the actual (i.e., physical) fields $F$ and $G$ as to their spectral plane-wave source fields $\cF$ and $\cG$, because of linearity for the superposition of Gaussian fields. For non-Gaussian $\cF$ and $\cG$, however, (\ref{eq:Isserlis}) does not necessarily extend to moments of $F$ and/or $G$.

\section{Power}
\subsection{1-D Cartesian Tangential or Normal Power Flow}
The complex Poynting vector 
$
\ul{S}
$ has Cartesian components $S_\alpha \ul{1}_\alpha = [( E_\beta H^*_\gamma - E_\gamma H^*_\beta ) / 2 ] \ul{1}_\alpha$ where $(\alpha, \beta, \gamma)$ is a cyclic permutation of $x$, $y$ and $z$. Their second-order moments are
\begin{align}
\langle |S_\alpha|^2 \rangle
&=
\frac{1}{4} \left ( \langle |E_\beta|^2 |H_\gamma|^2 \rangle 
+ 
\langle |E_\gamma|^2 |H_\beta|^2 \rangle \right . \nonumber\\
&~~~ \left. - \langle E_\beta H^*_\gamma E^*_\gamma H_\beta \rangle
 - \langle E^*_\beta H_\gamma E_\gamma H^*_\beta \rangle
 \right )
 \label{eq:Salpha_secondmoment}
\end{align}
and are evaluated using (\ref{eq:Isserlis}) for $\cF = \cE$ and $\cG = \cH$, as follows.

First, for a 1-D component of the power flowing parallel to the boundary, i.e., tangential $S_\alpha$ with $\alpha=x$ or $y$, Isserlis's theorem for complex $E_\beta$, $E_z$, $H_\beta$ and $H_z$ leads to
\begin{align}
\langle E_\beta H^*_z E^*_z H_\beta \rangle = 
\langle E^*_\beta H_z E_z H^*_\beta \rangle = 0
\label{eq:EHEH}
\end{align}
with $\beta = y$ or $x$, respectively ($\beta \not = \alpha$), because $\langle E_\beta H^{*}_z \rangle = \langle E_\beta E^{*}_z \rangle = \langle E_\beta H_\beta \rangle 
= 0$, as follows from (\ref{eq:EalphaHbeta_bis}) \cite{arnaLMAM}. 
The first two terms in (\ref{eq:Salpha_secondmoment}) with $\langle E_\beta H^{(*)}_z \rangle = \langle E_z H^{(*)}_\beta \rangle = 0$ are 
\begin{align}
\left \{ \begin{array}{l}
\langle |E_{\beta}(kz)|^2 |H_{z}(kz)|^2 \rangle\\
\langle |E_{z}(kz)|^2 |H_{\beta}(kz)|^2 \rangle
\end{array}
\right \}
&=
\frac{4 \, \langle |{\cal E}_0|^2 \rangle^2}{9 \, \eta^2_0} \left ( 1 \mp 2 j_0 \mp \frac{j_2}{2}  \right . \nonumber\\
&~~~ \left . 
+ j^2_0 - \frac{j^2_2}{2} + \frac{j_0 j_2}{2} \right )
\label{eq:avgEysqEzsq}
\end{align}
where upper and lower signs refer to $\langle |E_{\beta}|^2 |H_{z}|^2 \rangle$ and $\langle |E_{z}|^2 |H_{\beta}|^2 \rangle$. Substituting (\ref{eq:EHEH}) and (\ref{eq:avgEysqEzsq}) into (\ref{eq:Salpha_secondmoment}) yields
\begin{align}
\langle |S_{\alpha}(kz)|^2\rangle &=
\frac{2 \, \langle |{\cal E}_0|^2 \rangle^2}{9 \,  \eta^2_0} 
\left ( 1 + j^2_0 - \frac{j^2_2}{2} + \frac{j_0 j_2}{2} \right )
.
\label{eq:avgSxsq}
\end{align}
Combining (\ref{eq:avgSxsq}) with $\langle S_\alpha \rangle = 0$ results in the standard deviation $\sigma_{S_\alpha} = \sqrt{\langle |S_\alpha|^2 \rangle - | \langle S_\alpha \rangle |^2}$ as
\begin{align}
\sigma_{S_{\alpha}} (kz) = \sigma_{S_{\alpha,\infty}}
\sqrt{ 1 + j^2_0 - \frac{j^2_2}{2} + \frac{j_0 j_2}{2} },~~\alpha=x,y
\label{eq:sigma_Sx}
\end{align}
where its asymptotic value for $kz \rightarrow +\infty$ is
\begin{align}
\sigma_{S_{\alpha,\infty}} = \frac{\sqrt{2} \, \langle |{\cal E}_0|^2 \rangle}{3 \, \eta_0}.
\label{eq:sigma_Sx_infty}
\end{align}

Second, for the normally directed power flow ${S}_z$, we have
\begin{align}
\langle E_x H^*_y E^*_y H_x \rangle
&=
\langle E_x^* H_y E_y H^*_x \rangle
=
- \frac{\langle |{\cal E}_0|^2 \rangle^2}{\eta^2_0} j^2_1 
\label{eq:avgExHypEypHx}
\end{align}
on account of $\langle E_x H^*_y \rangle = - \langle E^*_y H_x \rangle = - \rmj \langle |\cE_0|^2 \rangle j_1/ \eta_0$ and $\langle E_x E^*_y \rangle = 0 = \langle E_x H_x \rangle$, the other terms in (\ref{eq:Salpha_secondmoment}) being
\begin{align}
&\/ \langle |E_x(kz)|^2 |H_y(kz)|^2 \rangle
= \langle |E_y(kz)|^2 |H_x(kz)|^2 \rangle = \nonumber \\
&\/ \frac{4 \, \langle |{\cal E}_0|^2 \rangle^2}{9 \, \eta^2_0}  
\left ( 1 - j^2_0 + \frac{9 j^2_1}{4}  - \frac{j^2_2}{4} + j_0 j_2 \right ) 
+ \frac{|\langle {\cal E}^2_0 \rangle|^2}{ \eta^2_0} j^2_1
.
\label{eq:avgExsqEysq_noncirc}
\end{align}
In order to arrive at (\ref{eq:avgExsqEysq_noncirc}), 
the pseudocovariances $\langle E_x H_y \rangle = - \langle E_y H_x \rangle = - \rmj \langle {\cE^2_0} \rangle  j_1 / \eta_0$ were employed. 
However, $\langle \cE^2_0\rangle = 0$ owing to the assumed complex circularity of the plane-wave fields, whence the second term in (\ref{eq:avgExsqEysq_noncirc}) vanishes.\footnote{In the evaluation of $\langle |E_x(kz)|^2 |H_y(kz)|^2 \rangle$, particularly for relatively small data sets and in case of (residual) noncircularity, a small value of the pseudovariance $\langle \cE^2_0\rangle$ may still have a significant effect when $\langle |E_x(kz)|^2 |H_y(kz)|^2 \rangle$ is also small, particularly when $kz \not \ll 1$.} 
Hence, substituting (\ref{eq:avgExHypEypHx}) and (\ref{eq:avgExsqEysq_noncirc}) into (\ref{eq:Salpha_secondmoment}) for $\alpha=z$ yields 
\begin{align}
\langle | S_z (kz) |^2 \rangle
&=
\frac{2 \, \langle |\cE_0|^2 \rangle^2}{9 \, \eta^2_0} 
\left ( 1 - j^2_0 + \frac{9 j^2_1}{2} - \frac{j^2_2}{4}  + j_0 j_2 \right ) 
\label{eq:avgSzsq}
\end{align}
which, combined with the mean (\ref{eq:avgSz}), results in
\begin{align}
\sigma_{S_z} (kz) = \sigma_{S_{z,\infty}} 
\sqrt{ 1 - j^2_0 - \frac{j^2_2}{4} + {j_0 j_2}}
\label{eq:sigma_Sz}
\end{align}
where 
\begin{align}
\sigma_{S_{z,\infty}} = \frac{\sqrt{2} \, \langle | \cE_0|^2 \rangle}{3 \, \eta_0}
.
\label{eq:sigma_Sz_infty}
\end{align}

\begin{figure}[htb] \begin{center}
\begin{tabular}{c}
\vspace{-0.5cm}\\
\hspace{-0.6cm}
\includegraphics[scale=0.49]{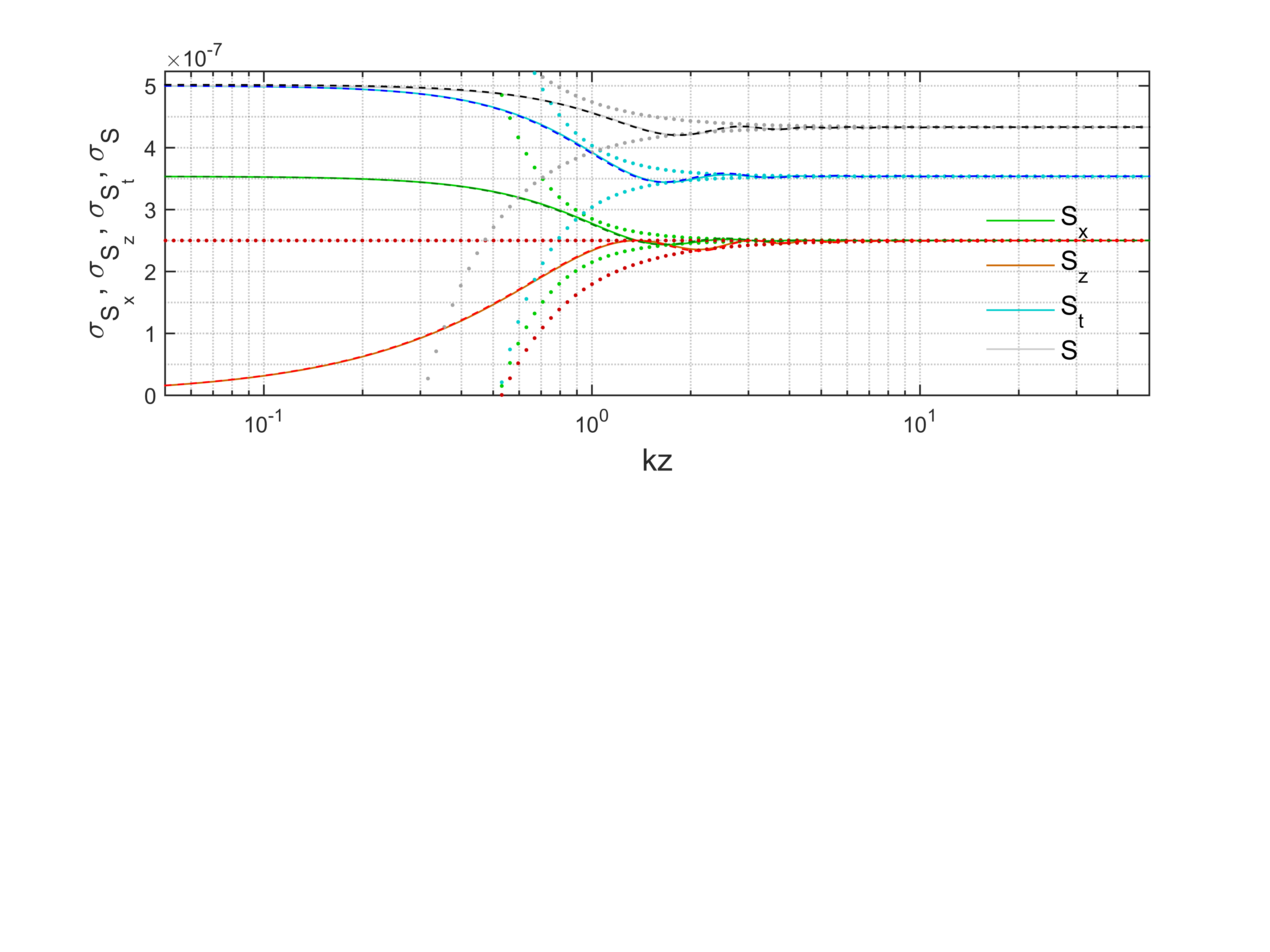}\\
\\
\vspace{-4.5cm}\\
(a)\\
\hspace{-0.8cm}
\includegraphics[scale=0.49]{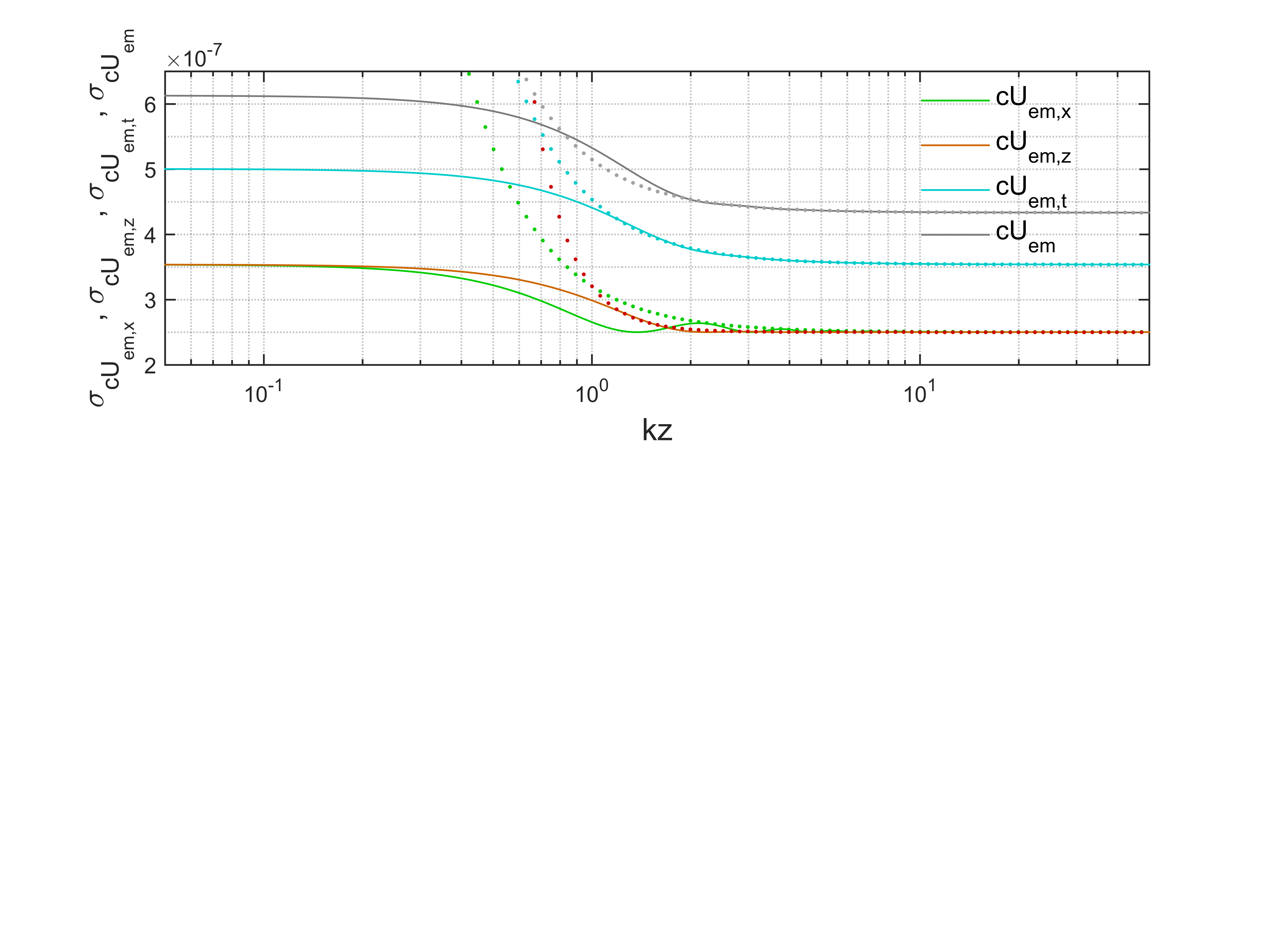}\\
\\
\vspace{-4.5cm}\\
(b)
\end{tabular}
\end{center}
{
\caption{\label{fig:sigmaSxSzStot}
\small
Standard deviations, in units W/m$^2$, based on $\langle {\cE^\prime_0}^2 \rangle^{1/2} = \langle {\cE^{\prime\prime}_0}^2 \rangle^{1/2} = 0.01$ V/m, for (a) Poynting power
$\sigma_{S_x}(kz)$, 
$\sigma_{S_z}(kz)$, 
$\sigma_{S_t}(kz)$, and  
$\sigma_{S}(kz)$ with upper and lower asymptotic envelopes $\Ups^{\pm}_{S_{(\alpha)}}(kz \gg 3/2)$. 
Solid: theory; 
dashed: MC simulation; 
dotted:  envelopes.
Olive/green: $S_x$;
maroon/red: $S_z$;
blue/cyan: $S_t$;
black/gray: $S$; and for
(b) scalar power
$\sigma_{\rmc U_{\rmem,x}}(kz)$, 
$\sigma_{\rmc U_{\rmem,z}}(kz)$, 
$\sigma_{\rmc U_{\rmem,t}}(kz)$, and
$\sigma_{\rmc U_{\rmem}}(kz)$
with upper asymptotic envelopes $\Ups^{+}_{\rmc U_{\rmem,\alpha}}(kz \gg 3/2)$, shown using the same color scheme as in (a).
}
}
\end{figure}

In Fig. \ref{fig:sigmaSxSzStot}(a), the standard deviations (\ref{eq:sigma_Sx}) and (\ref{eq:sigma_Sz}) are compared with numerical results obtained via Monte Carlo (MC) simulation of random plane waves, based on 30 random values of $|\cE_0|$, 32 elevation angles, 16 azimuthal angles, and $16$ polarization angles, evaluated at 362 logarithmically spaced distances $kz$ ranging from 0.05 to 50.
The value of $\sigma_{S_x}$ reaches its maximum on the PEC surface, where $\sigma_{S_z}$ vanishes. 
Local maxima and minima of $\sigma_{S_x} (kz)$ are found at distances that are marginally shorter than those for $\sigma_{S_z} (kz)$. The oscillations of $\sigma_{S_z} (kz)$ are more strongly damped than those for $\langle S_z (kz) \rangle$ \cite[Fig. 1]{arnaLMAM}, as will be confirmed in sec. \ref{sec:asymp} by comparing their envelopes.
For $kz \rightarrow +\infty$, it is  verified that
(\ref{eq:sigma_Sx}) and (\ref{eq:sigma_Sz}) merge, 
as expected for ideal isotropic fields in the absence of an EM boundary. 

\subsection{2-D Tangential Power Flow}
For the 2-D tangential Poynting vector $\ul{S}_t = S_x \ul{1}_x + S_y \ul{1}_y$, it follows that $|S_t|^2 \equiv \ul{S}^\dagger_t \cdot \ul{S}_t = |S_x|^2 + |S_y|^2$. With (\ref{eq:Isserlis}) and (\ref{eq:avgSx}), this yields $\sigma_{S_t} = \sqrt{2} \, \sigma_{S_x}$ as
\begin{align}
\sigma_{S_t}(kz) = \sigma_{S_{t,\infty}}
\sqrt{ 1 + j^2_0 - \frac{j^2_2}{2} + \frac{j_0 j_2}{2} }
\label{eq:sigma_St}
\end{align}
where 
\begin{align}
\sigma_{S_{t,\infty}} = \frac{2 \,  \langle | \cE_0|^2 \rangle }{ 3 \, \eta_0 } 
.
\label{eq:sigma_St_infty}
\end{align}

\subsection{3-D Spatial Power Flow}
For the full Poynting vector $\ul{S}=\sum_{\alpha=x,y,z}S_\alpha \ul{1}_\alpha$, we have $|S|^2 \equiv \ul{S}^\dagger \cdot \ul{S} = |S_t|^2 + |S_z|^2$. With (\ref{eq:Isserlis}), (\ref{eq:avgSxsq}) and (\ref{eq:avgSzsq}),
\begin{align}
\langle |S(kz)|^2 \rangle 
&=
\frac {2 \, \langle | \cE_0|^2 \rangle^2}{3 \, \eta^2_0} 
\left ( 1 + \frac{j^2_0}{3} + \frac{3 j^2_1}{2} - \frac{5 j^2_2}{12} + \frac{2 j_0 j_2}{3} \right ) 
\nonumber\\
\label{eq:avgSsq}
\end{align}
which, combined with (\ref{eq:avgSz}), yields
\begin{align}
\sigma_{S} (kz) = 
\sigma_{S_\infty} \sqrt{1 + \frac{j^2_0}{3} - \frac{5 j^2_2}{12} + \frac{2 j_0 j_2}{3}}
\label{eq:sigma_S}
\end{align}
where
\begin{align}
\sigma_{S_\infty} = \sqrt{\frac{2}{3}} \, \frac{\langle | \cE_0|^2 \rangle}{\eta_0}
.
\label{eq:sigma_S_infty}
\end{align}

\section{Energy\label{sec:energy}}
The standard deviations of $U_{\rme(,\alpha)}$ and $U_{\rmm(,\alpha)}$ for circular Gaussian fields near an impedance boundary were derived in \cite{arnaRS2007}, \cite{arnaTEMCmay2006}, enabling all moments to be obtained merely by a single variate transformation. 

For $|F_\alpha|^2$, setting $\cF_\alpha = \cF_\beta = \cG_\gamma = \cG_\delta$ in (\ref{eq:Isserlis}) and using $\langle \cF^2_\alpha \rangle = 0$ owing to the complex circularity ($\langle \cE^2_0 \rangle = 0$), this yields 
$
\langle |\cF_\alpha |^4 \rangle = 
2 \langle |\cF_\alpha|^2 \rangle^2 
+
|\langle \cF^2_\alpha \rangle|^2 
=
2 \langle |\cF_\alpha|^2 \rangle^2  
$,
leading to the familiar result 
$
\sigma_{|F_\alpha|^2} = 
\langle |F_\alpha|^2 \rangle
$ for $\alpha=x,y,z$
\cite{arnaTEMCmay2006}.
For the energies, this results in explicit functions of $kz$ as
\begin{align}
\sigma_{U_{\alpha}}(kz) &= \left ( 1 \mp j_0 \pm \frac{j_2}{2} \right ) \sigma_{U_{\alpha,\infty}},~~ \alpha=x,y
\label{eq:sigma_Ux}\\
\sigma_{U_{z}}(kz) &= \left ( 1 \pm j_0 \pm j_2 \right ) \sigma_{U_{\alpha,\infty}} 
\label{eq:sigma_Uz}
\end{align}
in which upper and lower signs refer to electric 
($U=U_{\rme}$, $\cF_0=\cE_0$, $\varphi_0=\eps_0$) and magnetic ($U=U_{\rmm}$, $\cF_0=\cH_0=\cE_0/\eta_0$, $\varphi_0=\mu_0$) energy densities, respectively, and with
\begin{align}
\sigma_{U_{\alpha,\infty}} = \langle U_{\alpha,\infty} \rangle = \frac{{\varphi_0 \langle |\cF_0|^2 \rangle}}{3},~~
\alpha = x,y,z.
\label{eq:sigma_Ualpha_infty}
\end{align}

For the energy density of the 2-D tangential electric or magnetic field, it follows from (\ref{eq:sigma_Ux}) and $U_t = U_x + U_y$ that
 \begin{align}
\sigma_{U_{t}} (kz)
&= \left ( 1 \mp j_0 \pm \frac{j_2}{2} \right ) \sigma_{U_{t,\infty}} 
\label{eq:sigma_Ut}
\end{align}
where
\begin{align}
\sigma_{U_{t,\infty}} = \frac{\sqrt{2}\, \varphi_0 \langle |\cF_0|^2 \rangle}{3}
.
\label{eq:sigma_Ut_infty}
\end{align}

For the full (vectorial) electric or magnetic energy density, the standard deviation follows from (\ref{eq:sigma_Ux}) and (\ref{eq:sigma_Uz}) as
\begin{align}
\sigma_{U}(kz) 
&= \sigma_{U_{\infty}}
\sqrt{
1 \mp \frac{2 j_0}{3} \pm \frac{4 j_2}{3} + j^2_0 + \frac{j^2_2}{2}
}
\label{eq:sigma_U}
\end{align}
where upper and lower signs again refer to electric and magnetic densities, respectively, and with
\begin{align}
\sigma_{U_{\infty}} = \frac{{\varphi_0 \langle |\cF_0|^2 \rangle}}{\sqrt{3}}
.
\label{eq:sigma_U_infty}
\end{align}
These standard deviations are plotted in Fig. \ref{fig:stdUe}, comparing favourably with results in Figs. 6(b) and 7(b) of \cite{arnaTEMCmay2006}.\footnote{Note that in \cite{arnaTEMCmay2006}, energy densities were denoted by the symbol $S$.}

For the combined EM energy density {$U_{\rmem(,\alpha)} \equiv U_{\rme(,\alpha)} + U_{\rmm(,\alpha)}$, the variances of its Cartesian tangential/normal, planar tangential, and full vectorial densities are found as
\begin{align}
\sigma^2_{U_{\rmem,\alpha}} &= \sigma^2_{U_{\rme,\alpha}} + \sigma^2_{U_{\rmm,\alpha}},~~\alpha=x,y,z \\
\sigma^2_{U_{\rmem(,t)}} &= \sigma^2_{U_{\rme(,t)}} + \sigma^2_{U_{\rmm(,t)}} + \left | \langle E_x H^*_y \rangle \right |^2 / (2\rmc)^2.
\end{align}
Explicit expressions for their standard deviations follow as
\begin{align}
\sigma_{U_{\rmem,\alpha}} (kz) &= \sigma_{U_{\rmem,\alpha,\infty}} \sqrt{1 + j^2_0 + \frac{j^2_2}{4} - j_0 j_2},~~ \alpha = x,y\label{eq:sigma_Uemx}
\\
\sigma_{U_{\rmem,z}} (kz) &= \sigma_{U_{\rmem,z,\infty}} \sqrt{1 + j^2_0 + j^2_2 + 2 j_0 j_2}\label{eq:sigma_Uemz}
\\
\sigma_{U_{\rmem,t}} (kz) &= \sigma_{U_{\rmem,t,\infty}} \sqrt{1 + j^2_0 + \frac{9 j^2_1}{4} + \frac{j^2_2}{4} - j_0 j_2}\label{eq:sigma_Uemt}
\\
\sigma_{U_{\rmem}} (kz) 
&= \sigma_{U_{\rmem,\infty}} \sqrt{
1 + j^2_0 + \frac{3j^2_1}{2} + \frac{j^2_2}{2} }
\label{eq:sigma_Uem}
\end{align}
where $\sigma_{U_{\rmem,(\alpha,)(t,)\infty}} = \sqrt{2} \, \sigma_{U_{(\alpha,)(t,)\infty}}$, i.e.,
\begin{align}
\sigma_{U_{\rmem,\alpha,\infty}} 
&= \frac{\sqrt{2} \, \varphi_0 \langle |\cF_0|^2 \rangle}{3} ,~~\alpha=x,y,z\label{eq:sigma_Uemalpha_infty}
\\
\sigma_{U_{\rmem,t,\infty}} 
&= \frac{2 \, \varphi_0 \langle |\cF_0|^2 \rangle}{3}
\label{eq:sigma_Uemt_infty}
\\
\sigma_{U_{\rmem,\infty}} 
&= \sqrt{\frac{2}{3}} \, \varphi_0 \langle |\cF_0|^2 \rangle
.
\label{eq:sigma_Uem_infty}
\end{align}
In (\ref{eq:sigma_Ux}), (\ref{eq:sigma_Uz}), (\ref{eq:sigma_Ut}) and (\ref{eq:sigma_U}), the terms linear in $j_0$ and $j_2$ cancel in their contributions to (\ref{eq:sigma_Uemx})--(\ref{eq:sigma_Uem}), leaving only quadratic terms. As shown in sec. \ref{sec:asymp}, this compensation between $U_{\rme(,\alpha)}$ and $U_{\rmm(,\alpha)}$ produces faster (higher-order) power-law decays with $kz$ in the envelopes of their sums $U_{\rmem(,\alpha)}$. 

In summary, a PEC boundary induces different RMS levels of fluctuation for the Poynting vs. scalar power, i.e.,
\begin{align}
\sigma_{S_{(\alpha)}}(kz) \not = \sigma_{\rmc U_{\rmem (,\alpha)}} (kz),~~\alpha=x,y,z,t
\label{eq:sigmaS_vs_cUem}
\end{align}
complementing the result $\rmj \langle S_{(\alpha)} (kz) \rangle  \not = \langle\rmc U_{\rmem(,\alpha)} (kz) \rangle$ \cite{arnaLMAM}.
Fig. \ref{fig:sigmaSxSzStot}(b) shows (\ref{eq:sigma_Uemx})--(\ref{eq:sigma_Uem_infty}), after scaling by $\rmc$.

\begin{figure}[htb] \begin{center}
\begin{tabular}{c}
\hspace{-0.6cm}
\includegraphics[scale=0.49]{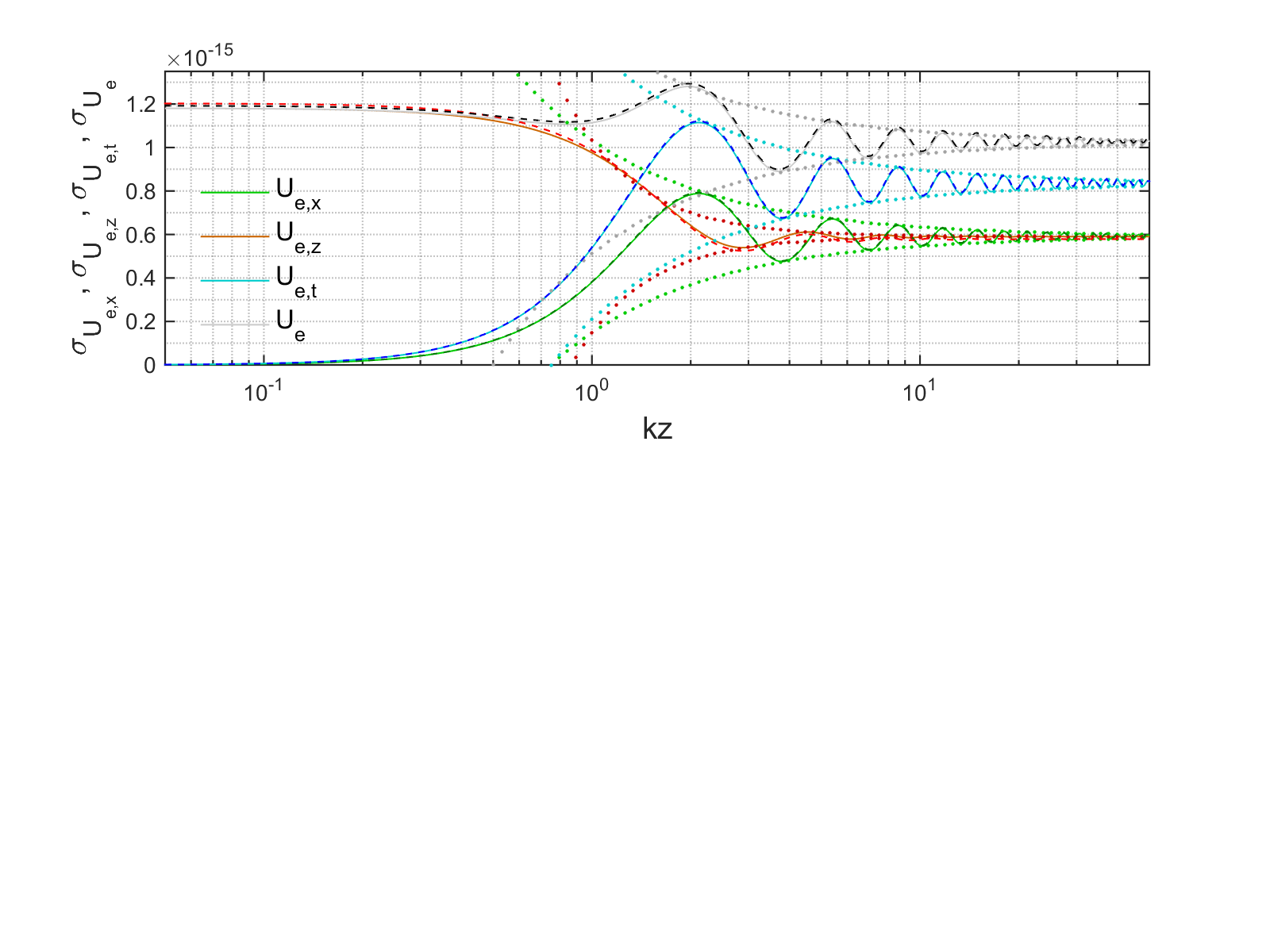}\\
\vspace{-4cm}\\
(a)\\
\hspace{-0.6cm}
\includegraphics[scale=0.49]{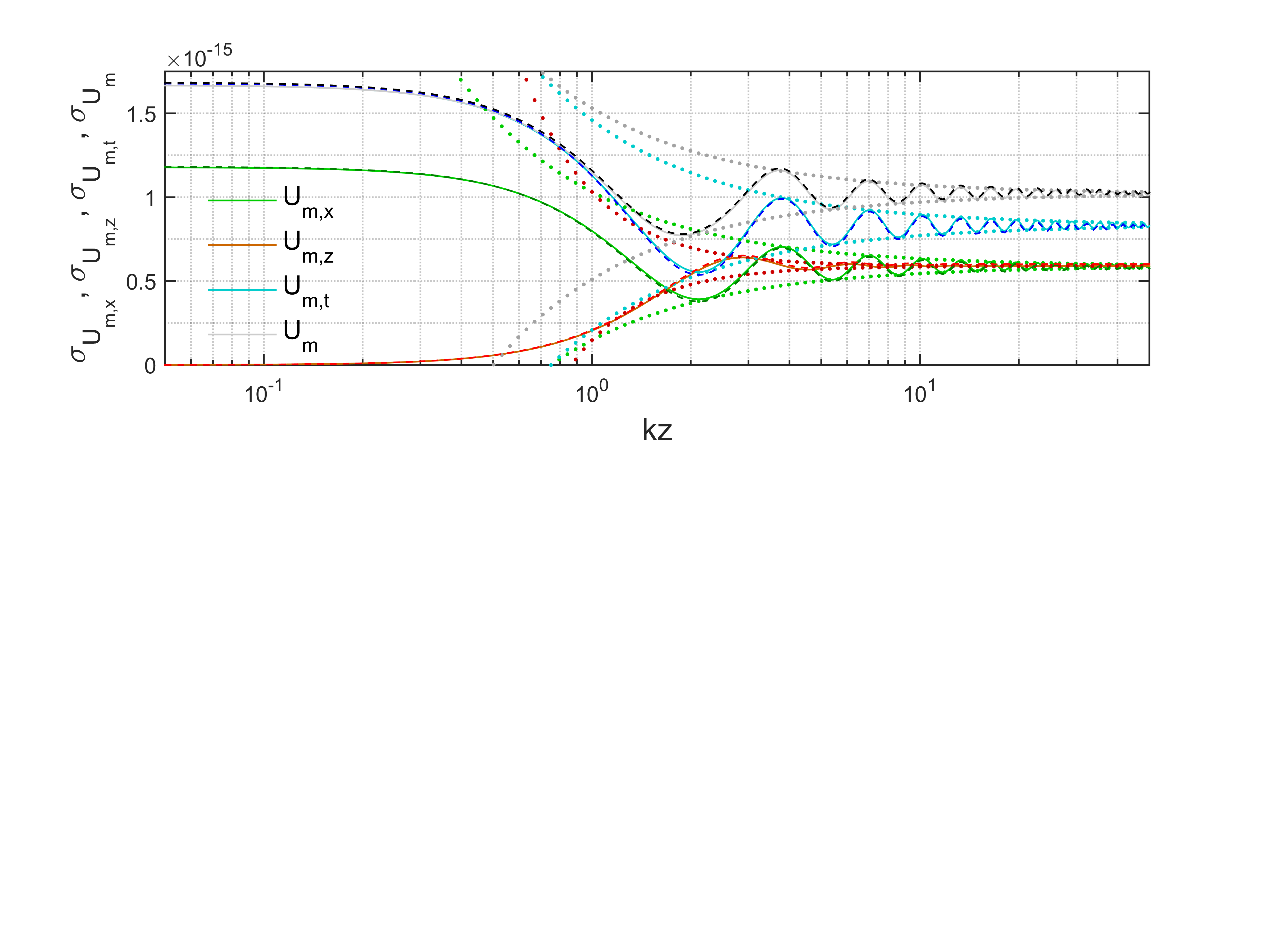}\\
\vspace{-4cm}\\
(b)
\end{tabular}
\end{center}
{
\caption{\label{fig:stdUe}
\small
Standard deviations, in units J/m$^3$, based on $\langle {\cE^\prime_0}^2 \rangle^{1/2} = \langle {\cE^{\prime\prime}_0}^2 \rangle^{1/2} = 0.01$ V/m, for 
(a) $U_{\rme,x}$, $U_{\rme,z}$, $U_{\rme,t}$, $U_{\rme}$, and (b) $U_{\rmm,x}$, $U_{\rmm,z}$, $U_{\rmm,t}$, $U_{\rmm}$.
Solid: theory; 
dashed: MC simulation; 
dotted: asymptotic upper and lower envelopes.
Olive/green: $U_{(\rme,)(\rmm,)x}$;
maroon/red: $U_{(\rme,)(\rmm,)z}$;
blue/cyan: $U_{(\rme,)(\rmm,)t}$;
black/gray: $U_{(\rme)(\rmm)}$.
}
}
\end{figure}

In applications to immunity, susceptibility or fading testing, the PEC boundary thus induces a $kz$-dependent scaling of the probability density functions (PDFs). For example, for the Cartesian field strength $|E_{\alpha}|$ with variance $\sigma^2_{|E_\alpha|}(kz) = [(4-\pi)/\eps_0] \sigma_{U_{\rme,\alpha}}(kz)$ for $\alpha=x,y,z$, the scaling of its PDF
\begin{align}
f_{|E_\alpha|}(|e_\alpha|,kz) = 
\frac{\eps_0\,|e_\alpha| }{2 \sigma_{U_{\rme,\alpha}}(kz)} \exp \left [ - \frac{\eps_0\,|e_\alpha|^2}{4 \sigma_{U_{\rme,\alpha}}(kz)} \right ]
\label{eq:Rayleigh_kzscaled}
\end{align}
is damped oscillatory, as governed by (\ref{eq:sigma_Ux}) or (\ref{eq:sigma_Uz}). In turn, this scaling affects associated statistics and distributions, e.g.,  
$F_{|E_\alpha|_{\rm max}}(|e_\alpha|_{\rm max},kz) = [ F_{|E_\alpha|}(|e_\alpha|_{\rm max},kz)
]^{N(kz)} $ 
for the maximum field strength $|E_\alpha|_{\rm max}$.
Fig. \ref{fig:PDF_magEx_magEz} demonstrates this scaling and bifurcation of (\ref{eq:Rayleigh_kzscaled}) for actual (i.e., dimensioned, non-normalized) $|E_x|$ vs. $|E_z|$, at selected values of $kz$.

\begin{figure}[htb] \begin{center}
\begin{tabular}{c}
\vspace{-0.5cm}\\
\hspace{-0.6cm}
\includegraphics[scale=0.49]{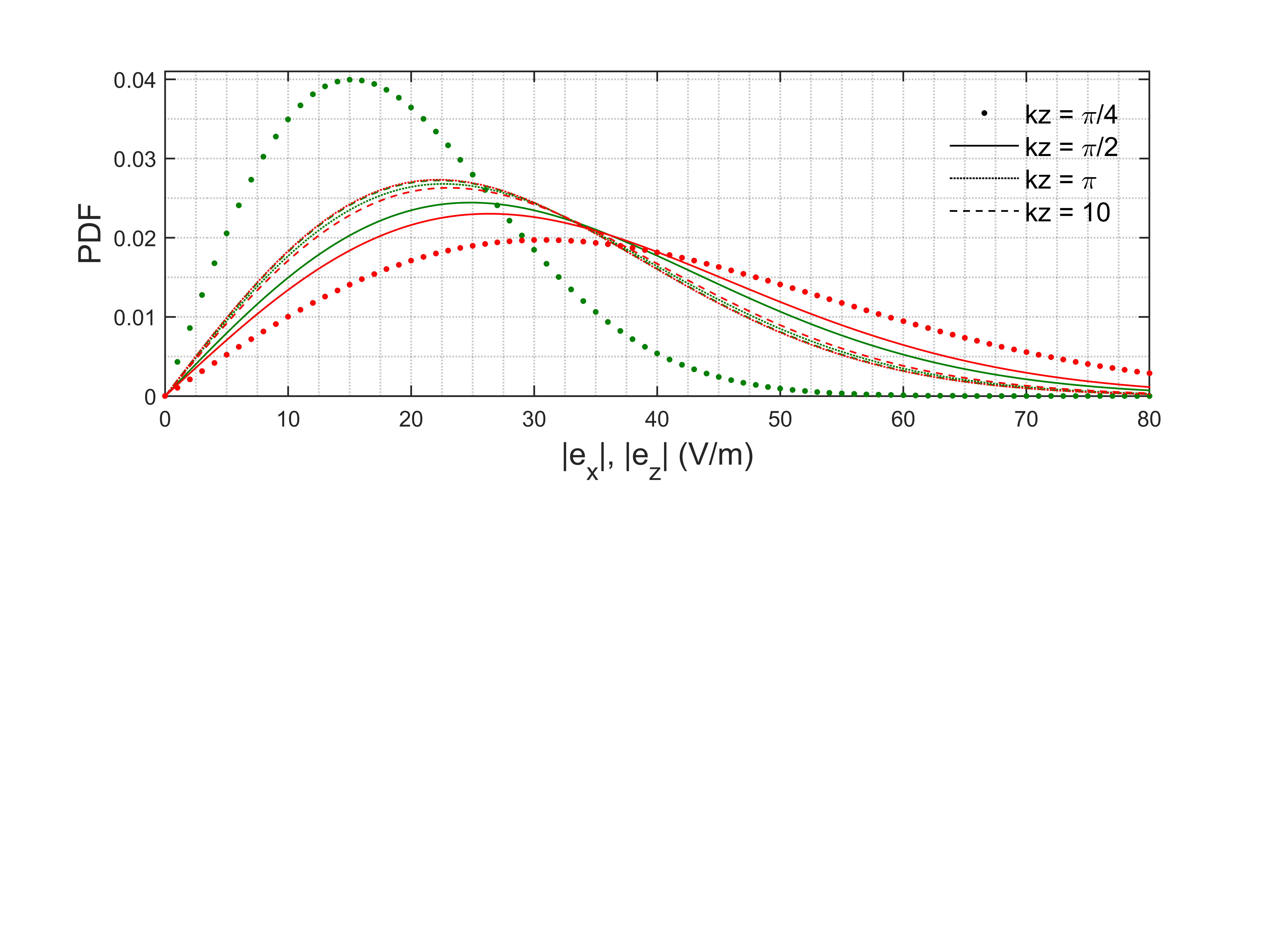}\\
\vspace{-4.5cm}\\
\end{tabular}
\end{center}
{
\caption{\label{fig:PDF_magEx_magEz}
\small
Rayleigh PDFs (\ref{eq:Rayleigh_kzscaled}) of $|E_x|$ (green) and $|E_z|$ (red) for $\langle {\cE^\prime_0}^2 \rangle^{1/2} = \langle {\cE^{\prime\prime}_0}^2 \rangle^{1/2} = 20$ V/m, exhibiting $kz$-dependent scaling.
}
}
\end{figure}

\section{Asymptotic Envelopes and Range of $\sigma$\label{sec:asymp}}
In \cite[sec. III-C]{arnaLMAM}, asymptotic approximations by envelopes for the mean power and energy densities at $kz \gg 3/2$ were derived, resulting in simpler inverse power law expressions for their damped oscillations. 
Such nonlocal approximations are also useful when phase shifts arise that are caused by perturbations, e.g., apertures, scatterers, other nearby walls, boundary roughness, antenna-surface coupling, etc. Here, results for these envelopes of the mean are extended to corresponding envelopes for the standard deviation. 

\subsection{Linear Forms for Electric or Magnetic Energy Fluctuations\label{sec:linearform}}
In (\ref{eq:sigma_Ux}),
(\ref{eq:sigma_Uz}) and
(\ref{eq:sigma_Ut}), the standard deviations have the form 
\begin{align}
\sigma_{U_\alpha}(kz) = (1 + a_0 j_0 + a_2 j_2) \sigma_{U_{\alpha,\infty}} .
\label{eq:sdUalpha_gen}
\end{align} 
Their associated asymptotic envelopes $\Ups^{\pm}_{U_\alpha}(kz \gg 3/2)$ can be obtained as the modulus of the corresponding analytic signal, $|\tilde{\sigma}_{U_\alpha} (kz \gg 3/2)|$. 
To this end, $\tilde{j}_\ell \stackrel{\Delta}{=} j_\ell + \rmj {\rmH}[j_\ell]$ for the individual terms in (\ref{eq:sdUalpha_gen}) are sought for $kz \gg 3/2$, where $\rmH[\cdot]$ denotes the Hilbert transform.
Recall the asymptotic approximations 
\begin{align}
j_\ell &\simeq \frac{\sin(2kz-\ell\pi/2)}{2kz}~~{\rm for}~kz\gg \frac{\ell(\ell+1)}{4}
\label{eq:asympBessel_HF}
\end{align}
which, for even $\ell \equiv 2m$, result in
\begin{align}
&\, \rmH [ j_{2m} (2kz \gg (2m+1)m)] \simeq (-1)^m \, \frac{1-\cos(2kz)}{2kz}
.
\label{eq:conj_asympBessel_HF}
\end{align}
On the other hand, $z^{-1} \sin(2kz-\ell\pi/2)$ in (\ref{eq:asympBessel_HF}) represents an amplitude modulated signal when $z \gg (2k)^{-1}$, for an arbitrary wavenumber $k>0$. Therefore, on application of Bedrosian's theorem \cite{king2009}, it follows that
\begin{align}
\tilde{j}_\ell = j_\ell + \rmj {\rmH}[j_\ell] 
\simeq \frac{\exp[\rmj (2 kz - (\ell + 1) \pi/2 ) ]}{ 2kz }
\end{align}
whence $|\tilde{j}_\ell(2kz \gg 1)| \simeq (2kz)^{-1}$.
Combining with $\rmH[1]=0$ and applying the linearity of the Hilbert transform, upper (`$+$') and lower (`$-$') envelopes of $\sigma_{U_\alpha}(kz \gg 3/2)$ follow as 
\begin{align}
\Ups^\pm_{U_\alpha} (kz \gg 3/2)  
\simeq 
\left \{
\begin{array}{ll}
\left [ 1 \pm \frac{|a_0-a_2|}{2(kz)} \right ] \sigma_{U_{\alpha,\infty}},& a_0 - a_2 \not = 0,\\
\\
\left [ 1 \pm \frac{3|a_2|}{4(kz)^2} \right ] \sigma_{U_{\alpha,\infty}},& a_0 -  a_2 = 0.
\end{array}
\right .
\label{eq:env_Ualpha_lin}
\end{align}
Thus, the envelopes decay typically\footnote{
If $a_0 - a_2 = 0$, e.g., for $U_z$, the envelopes are found by first replacing $j_2$ by its exact expression $[3/(2kz)] j_1 - j_0$ instead of the approximation $-j_0$.
} according to a first-order power law, $(kz)^{-1}$.
Application of (\ref{eq:env_Ualpha_lin}) to (\ref{eq:sigma_Ux}),
(\ref{eq:sigma_Uz}) and
(\ref{eq:sigma_Ut}) yields the asymptotic envelopes of $\sigma_{U_{(\alpha)(t)}}(kz \gg 3/2)$ as 
\begin{align}
\Ups^\pm_{U_\alpha}  (kz \gg 3/2)
&\simeq
\left [ 1 \pm \frac{3}{4kz} \right ] \sigma_{U_{\alpha,\infty}},~~\alpha=x,y 
\label{eq:sigma_Ux_env}\\
\Ups^\pm_{U_z} (kz \gg 3/2)
&\simeq
\left [ 1 \pm \frac{3}{4(kz)^2} \right ] \sigma_{U_{z,\infty}} 
\label{eq:sigma_Uz_env}\\
\Ups^\pm_{U_t} (kz \gg 3/2)
&\simeq
\left [ 1 \pm \frac{3}{4kz} \right ] \sigma_{U_{t,\infty}} 
.
\label{eq:sigma_Ut_env}
\end{align}

By contrast, close to the PEC boundary, using 
\begin{align}
j_\ell \simeq \frac{(2kz)^{\ell}}{(2\ell+1)!!} \left [ 1 - \frac{(2kz)^2}{2(2\ell+3)} \right ] ~~ {\rm for}~kz \ll \sqrt{2\ell+5}
\end{align}
all $\sigma_{U_{(\alpha)(t)}} (kz)$ are quadratic in $kz$ to leading order, viz.,
\begin{align}
\sigma_{U_{\rme(,\alpha)(,t)}} (kz \ll \sqrt{5}) 
&\simeq
\frac{4(kz)^2}{5}
\sigma_{U_{\rme(,\alpha)(,t),\infty}},~~\alpha=x,y \label{eq:sigmaUext_smallkz}\\
\sigma_{U_{\rme,z}} (kz \ll \sqrt{5}) 
&\simeq
2 \left [ 1 - \frac{(kz)^2}{5} \right ]\sigma_{U_{\rme,z,\infty}} \label{eq:sigmaUez_smallkz}\\
\sigma_{U_{\rmm(,\alpha)(,t)}} (kz \ll \sqrt{5}) 
&\simeq
2 \left [ 1 - \frac{2(kz)^2}{15} \right ]\sigma_{U_{\rmm(,\alpha)(,t),\infty}} 
\label{eq:sigmaUmzt_smallkz}\\
\sigma_{U_{\rmm,z}} (kz \ll \sqrt{5}) 
&\simeq
\frac{2(kz)^2}{5}\sigma_{U_{\rmm,z,\infty}}  \label{eq:sigmaUmz_smallkz}
.
\end{align}

\subsection{Quadratic Forms for EM Energy or Power Fluctuations\label{sec:quadraticform}}
In (\ref{eq:sigma_Sx}),
(\ref{eq:sigma_Sz}), 
(\ref{eq:sigma_St}) and
(\ref{eq:sigma_S}), the variances have the form 
\begin{align}
\sigma^2_{S_{(\alpha)}}(kz) = (1 + a_{00} j^2_0 + a_{22} j^2_2 + a_{02} j_0 j_2) \sigma^2_{S_{(\alpha,)\infty}} 
.
\label{eq:varSalpha_gen}
\end{align} 
For their asymptotic envelopes, a derivation analogous to that in sec. \ref{sec:linearform} can be performed. With
\begin{align}
j^2_{2m} \simeq \frac{2 \left [ 1 - \cos(4kz) \right ]}{ (4kz)^2} &, \,
\rmH [ j^2_{2m}] \simeq \frac{2 \left [ 4kz - \sin(4kz)\right ]}{(4kz)^2} \nonumber\\ 
j_{2m} j_{2m+2} \simeq \frac{2\sin(4kz)}{ (4kz)^2} &, \,
\rmH [ j_{2m} j_{2m+2}] \simeq - \rmH [ j^2_{2m}] 
\label{eq:conj_asympBesselsqmixed_HF}
\end{align}
for $kz \gg 3/2$, envelopes for $\sigma^2_{S_{(\alpha)}}(kz)$ can then be converted to envelopes $\Ups^\pm_{S_{(\alpha)}} (kz)$ for $\sigma_{S_{(\alpha)}}(kz)$, on expanding $\sigma_{S_{(\alpha)}}(kz)$ in (\ref{eq:varSalpha_gen}) using $\sqrt{1+x} \simeq 1 + {x}/{2}$ for $x \ll 1$. 
This results in\footnote{A similar remark as in sec. \ref{sec:linearform} for $U_z$ holds if $a_{00} + a_{22} - a_{02}=0$.} 
\begin{align}
&\Ups^\pm_{S_{(\alpha)}} (kz \gg 3/2) \simeq \sigma_{S_{(\alpha,)\infty}} \nonumber\\
&\, \times
\left \{
\begin{array}{ll} 
\left [ 1 \pm \frac{|a_{00} + a_{22} - a_{02}|}{8(kz)^2} \right ]
,& a_{00}+a_{22}-a_{02} \not = 0,\\
\\ 
\left [ 1 \pm \frac{3|2a_{22}-a_{02}|}{32 (kz)^3} \right ] ,& a_{00}+a_{22}-a_{02} = 0.
\end{array}
\right .
\label{eq:asympenv_S_largekz}
\end{align}
The fluctuations of the tangential power are decaying at a faster (cubic) rate than the quadratic rate of the normal and full vector Poynting powers, viz.,
\begin{align}
&\Ups^\pm_{S_{(\alpha)(t)}} (kz \gg 3/2)
\simeq
\left [ 1 \pm \frac{9}{64(kz)^3} \right ] \sigma_{S_{(\alpha,)(t,)\infty}} 
, ~~ \alpha=x,y
\label{eq:sigma_Sx_env}\\
&\Ups^+_{S_z}
=
\sigma_{S_{z,\infty}},
~
\Ups^-_{S_z} (kz \gg 3/2)
\simeq 
\left [ 1 - \frac{9}{32(kz)^2}\right ] \sigma_{S_{z,\infty}} 
\label{eq:sigma_Sz_env}\\
&\Ups^\pm_{S} (kz \gg 3/2)
\simeq 
\left [ 1 \pm \frac{3}{32(kz)^2}\right ] \sigma_{S_{\infty}} 
.
\label{eq:sigma_S_env}
\end{align}
Unlike $\Ups^\pm_{U_{(\alpha)}} (kz)$, which exhibit the same $kz$-dependence as the envelopes $\Xi^\pm_{U_{(\alpha)}} (kz)$ for $\langle U_{(\alpha)} (kz) \rangle$ \cite[eqs. (34)--(36)]{arnaLMAM}, the $\Ups^\pm_{S_z}(kz)$ decay more rapidly than ${\mit \Xi}^\pm_{S_z}(kz)$ \cite[eq. (33)]{arnaLMAM}. 
Near the boundary, $\sigma_{S_z}(kz)$ exhibits a {\em linear\/} dependence, all other Poynting powers being quadratic in $kz$, i.e., 
\begin{align}
\sigma_{S_{(\alpha)(t)}} (kz \ll \sqrt{5}) 
&\simeq
\sqrt{2} \left [ 1 - \frac{3(kz)^2}{10} \right ]\sigma_{S_{(\alpha,)(t,)\infty}},~\alpha=x,y \label{eq:sigmaSxt_smallkz}\\
\sigma_{S_{z}} (kz \ll \sqrt{5}) 
&\simeq
\sqrt{\frac{8}{5}}\,kz\,\sigma_{S_{z,\infty}} \label{eq:sigmaSz_smallkz}\\
\sigma_{S} (kz \ll \sqrt{5}) 
&\simeq
\frac{2}{\sqrt{3}} \left [ 1 - \frac{(kz)^2}{10} \right ] \sigma_{S_{\infty}}
. \label{eq:sigmaS_smallkz}
\end{align}

For the total EM energy densities (\ref{eq:sigma_Uemx})--(\ref{eq:sigma_Uem}), 
the upper envelopes $\Ups^{+}_{U_{\rmem(,\alpha)(,t)}} (kz)$ 
follow in a similar way as
\begin{align}
\Ups^+_{U_{\rmem,\alpha}} (kz \gg 3/2)
&\simeq 
\left [ 1 + \frac{9}{32(kz)^2}\right ] \sigma_{U_{\rmem,\alpha, \infty}} 
,~\alpha=x,y
\label{eq:sigma_Uemx_env}\\
\Ups^+_{U_{\rmem,z}} (kz \gg 3/2)
&\simeq 
\left [ 1 + \frac{9}{32(kz)^4}\right ] \sigma_{U_{\rmem,z, \infty}} 
\label{eq:sigma_Uemz_env}\\
\Ups^+_{U_{\rmem,t}} (kz \gg 3/2)
&\simeq 
\left [ 1 + \frac{9}{32(kz)^2}\right ] \sigma_{U_{\rmem,t, \infty}} 
\label{eq:sigma_Uemt_env}\\
\Ups^+_{U_{\rmem}} (kz \gg 3/2)
&\simeq 
\left [ 1 + \frac{3}{16(kz)^2}\right ] \sigma_{U_{\rmem,\infty}} 
.
\label{eq:sigma_Uem_env}
\end{align}
In particular, $\Ups^+_{S} (kz \gg 3/2)$ and $\Ups^+_{\rmc U_{\rmem}} (kz \gg 3/2)$ decay at identical quadratic rates and magnitudes. 
The $\Ups^{-}_{U_{\rmem,\alpha}} (kz \gg 3/2)$ coincide with $\sigma_{U_{\rmem,\alpha,\infty}}$, whereas $\Ups^{-}_{{U_{\rmem,t}}}(kz \gg 3/2)$ and $\Ups^{-}_{{U_{\rmem}}}(kz \gg 3/2)$ exhibit no significant oscillations, whence
\begin{align}
\Ups^-_{U_{\rmem,\alpha}} (kz \gg 3/2)
&\simeq 
\sigma_{U_{\rmem,\alpha,\infty}},~~\alpha=x,y,z \label{eq:sigma_Uemx_env_min}\\
\Ups^-_{U_{\rmem(,t)}} (kz \gg 3/2) 
&\simeq \Ups^+_{U_{\rmem(,t)}} (kz \gg 3/2)
.
\label{eq:sigma_Uem_env_min}
\end{align}

Near the boundary, $\sigma_{\rmc U_{\rmem{(,x)(,t)}}}(kz) \simeq \sigma_{S_{(x)(t)}} (kz)$ while $\sigma_{\rmc U_{\rmem{(,z)}}}(kz) > \sigma_{S_{(z)}} (kz)$, as seen in Fig. \ref{fig:sigmaSxSzStot}. Explicitly,
\begin{align}
\sigma_{U_{\rmem,\alpha}} (kz \ll \sqrt{5}) 
&\simeq
\sqrt{2} \left [ 1 - \frac{2(kz)^2}{5} \right ] \sigma_{U_{\rmem,\alpha,\infty}},~\alpha=x,y \label{eq:sigmaUemx_smallkz}\\
\sigma_{U_{\rmem,z}} (kz \ll \sqrt{5}) 
&\simeq
\sqrt{2} \left [ 1 - \frac{(kz)^2}{5} \right ] \sigma_{U_{\rmem,z,\infty}} \label{eq:sigmaUemz_smallkz}\\
\sigma_{U_{\rmem,t}} (kz \ll \sqrt{5}) 
&\simeq
\sqrt{2} \left [ 1 - \frac{3(kz)^2}{10} \right ] \sigma_{U_{\rmem,t,\infty}} \label{eq:sigmaUemt_smallkz}\\
\sigma_{U_\rmem} (kz \ll \sqrt{5}) 
&\simeq
\sqrt{2} \left [ 1 - \frac{(kz)^2}{2} \right ] \sigma_{U_{\rmem,\infty}} \label{eq:sigmaUem_smallkz}
.
\end{align}

\subsection{Mixed Linear-Quadratic Forms for 3-D Vectorial Electric or Magnetic Energy Fluctuations\label{sec:linearquadraticform}}
If $\sigma(kz)$ contains both linear and quadratic terms in $j_\ell$, e.g., for $U$ in (\ref{eq:sigma_U}), then the weaker decay for the linear terms -- typically a $(kz)^{-1}$-law -- prevails for $kz \gg 3/2$. Thus,
\begin{align}
\Ups^\pm_U  (kz \gg 3/2)
&\simeq
\left [ 1 \pm \frac{1}{2kz}\right ]
\sigma_{U_{\infty}} 
\label{eq:asympenv_U_largekz}
\end{align}
for $U=U_{\rme}$ or $U_{\rmm}$.
Near the boundary,
\begin{align}
\sigma_{U_{\rme}} (kz \ll \sqrt{5})
&\simeq
\frac{2}{\sqrt{3}} \left [ 1 - \frac{2(kz)^2}{15} \right ] \sigma_{U_{\rme,\infty}} \label{eq:sigmaUe_smallkz}\\
\sigma_{U_{\rmm}} (kz \ll \sqrt{5}) 
&\simeq
\sqrt{\frac{8}{3}} \left [ 1 - \frac{2(kz)^2}{3} \right ]\sigma_{U_{\rmm,\infty}}. \label{eq:sigmaUm_smallkz}
\end{align}

\section{Relative Levels of Fluctuation}
\subsection{Coefficient of Variation}
The coefficient of variation (CV), $\nu_X\stackrel{\Delta}{=}\sigma_X / \langle X \rangle$, represents the mean-normalized RMS level of fluctuation of $X$, as a simple measure of its relative uncertainty. Its reciprocal, $1 / \nu_X$, measures the SNR of $X$. 
Close to the boundary, the CV differs considerably between the various components of the powers $S$ or $\rmc U_{\rmem}$. On the boundary itself,
\begin{align}
&\, \nu_{S_z}(0) = \rmj {2}/{\sqrt{5}},~~\nu_{S}(0) = \lim_{kz \rightarrow 0+}
\rmj {\sqrt{2}}/(kz) = + \rmj \infty\label{eq:nu_Sz_smallkz}\\
&\, \nu_{\rmc U_{\rmem,\alpha}}(0) = \sqrt{2}\, \nu_{\rmc U_{\rmem,t}}(0) = \sqrt{3}\, \nu_{\rmc U_{\rmem}}(0) = 1.
\label{eq:nu_Uem_smallkz}
\end{align}
Fig. \ref{fig:nu}(a) shows the SNRs $\rmj/ \nu_{S_{z}}(kz)$ and $\rmj/ \nu_{S}(kz)$, confirming that $|\nu_{S}(kz)| \geq |\nu_{S_z}(kz)|$ for arbitrary $kz$ and that both CVs are in phase for $kz \gg 3/2$. 
Fig. \ref{fig:nu}(b) shows selected CVs $\nu_{U_{(\rmem)(\alpha)}}(kz) \equiv \nu_{\rmc U_{(\rmem)(\alpha)}}(kz)$.
Comparing with Fig. \ref{fig:nu}(a), a slightly smaller relative uncertainty of $S_z$ over $U_{\rme,z}$ (up to $11.8\%$, on the boundary) is achieved for $kz < 1$, while for $kz > 1$ the relative uncertainty for $S_z$ is much larger. The constancy of $\nu_{U_{\rme,z}}(kz)$ and $\nu_{U_{\rmm,t}}(kz)$ indicates that the $\chi^2_2$ and $\chi^2_4$ PDFs of the respective deep energies are maintained at any distance, up to the boundary. The variation of $\nu_{U_\rme}(kz)$ from $1$ at $kz=0$ to $1/\sqrt{3}$ at $kz \rightarrow +\infty$ is a testimony of the compound exponential PDF \cite{arnaCE} for the full $U_\rme(kz)$, evolving from $\chi^2_2$ to $\chi^2_6$. The $U_{\rmem}(kz)$ has a corresponding evolution, from $\chi^2_6$ for $U_{\rmem}(0)=U_{\rme,z}(0)+U_{\rmm,t}(0)$ with $\nu_{U_{\rmem}}(0) = 1/\sqrt{3}$, to $\chi^2_{12}$ for $U_{\rme}(kz\rightarrow\infty) + U_{\rmm}(kz\rightarrow\infty)$ and $\nu_{U_{\rmem}}(\infty) = 1/\sqrt{6}$.
All $\nu_{\rmc U_{\rmem(,\alpha)}}(kz)$ are rescalings of the
$\sigma_{\rmc U_{\rmem(,\alpha)}}(kz) \equiv \rmc \, \sigma_{U_{\rmem(,\alpha)}}(kz)$ shown in Fig. \ref{fig:sigmaSxSzStot}(b), because all $\langle U_{\rmem(,\alpha)} \rangle$ are independent of $kz$.
Fig. \ref{fig:nu} also indicates that  
\begin{align}
\nu_{\rmc U_{\rmem}}(kz) \ll |\nu_{S}(kz)|
\label{eq:nu_cUem_vs_S}
\end{align} 
at any $kz$, differing typically by an order of magnitude or more. As a practical consequence, the conversion of the intensity or energy measured using electric and magnetic field probes to the scalar EM power produces considerably lower relative uncertainty than the direct measurement of Poynting power using aperture antennas, at any distance from the boundary.

\begin{figure}[htb] \begin{center}
\begin{tabular}{c}
\vspace{-0.5cm}\\
\hspace{-0.6cm}
\includegraphics[scale=0.49]{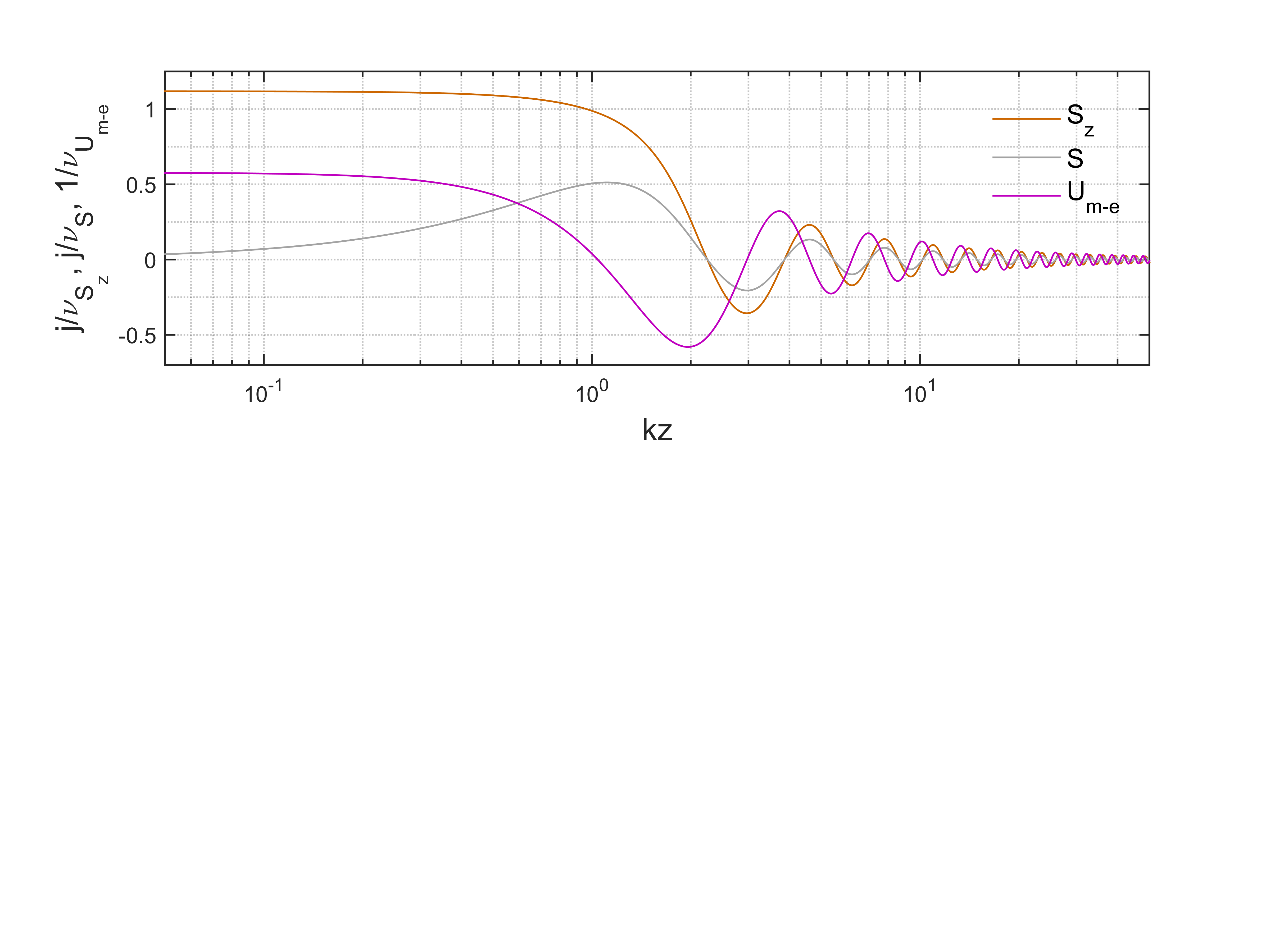}\\
\vspace{-4cm}\\
(a)\\
\hspace{-0.6cm}
\includegraphics[scale=0.49]{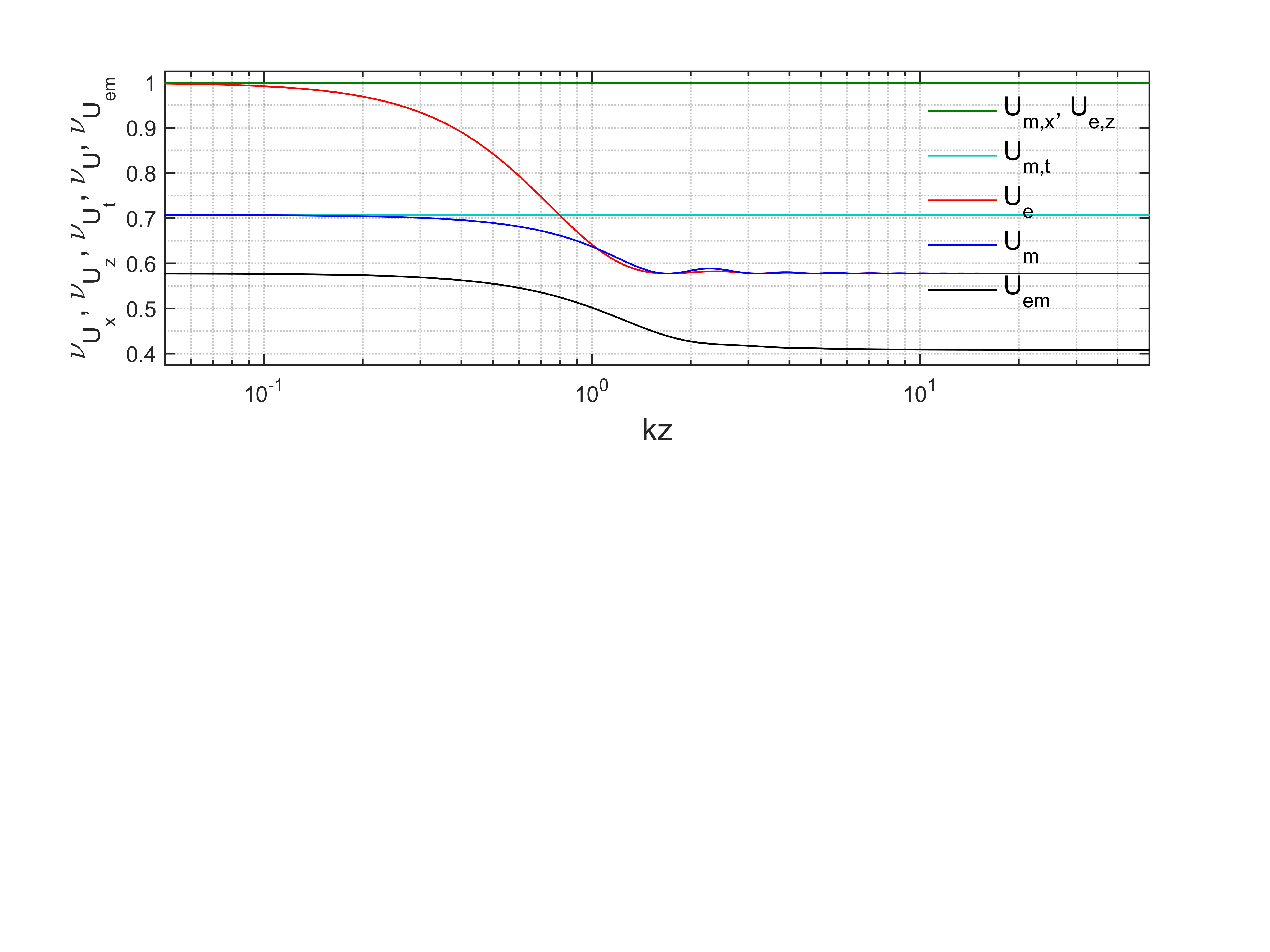}\\
\vspace{-4cm}\\
(b)
\end{tabular}
\end{center}
{
\caption{\label{fig:nu}
\small
Relative fluctuations as a function of $kz$: 
(a) SNRs $\rmj/\nu_{S_z}$ (brown), $\rmj/\nu_{S}$ (gray), and $1/ \nu_{U_{\rmm-\rme}}$ (purple);
(b) CVs $\nu_{U_{\rmm,x}}$ (green), $\nu_{U_{\rme,z}}$ (green), $\nu_{U_{\rmm,t}}$ (cyan), $\nu_{U_\rme}$ (red), $\nu_{U_\rmm}$ (blue), and $\nu_{U_\rmem}$ (black). 
}
}
\end{figure}

\subsection{Coefficient of Range Variation}
The CV measures the actual standard deviation relative to the actual mean value, both at the same location.
Related to their respective envelopes,
the {\it coefficient of range variation} (CRV) of $X$, denoted as $\zeta_X$, is here introduced as the ratio of the local range of variation between $\Ups^+_X(kz)$ and $\Ups^-_X(kz)$ for $\sigma_X$ at $kz$ to the corresponding range of ${\mit \Xi}^{\pm}_X(kz)$ for the real or complex $\langle X \rangle$ at the same  $kz$, i.e.,
\begin{align}
\zeta_X(kz) \stackrel{\Delta}{=} \frac{\Ups^+_X(kz) - \Ups^-_X(kz)}{{\mit \Xi}^+_X(kz) - {\mit \Xi}^-_X(kz)}
.
\end{align}
This measure is particularly useful when nonlocal deviations are expected from ideal $\sigma_X(kz)$ or $\langle X(kz) \rangle$ and hence $\nu_X(kz)$, e.g., for causes of field perturbations listed in sec. \ref{sec:asymp}.
For $kz \gg 3/2$, the CRVs for the Poynting power components are
\begin{align}
\zeta_{S_\alpha} (kz \gg 3/2) 
&= \frac{9 \, \sigma_{S_{\alpha,\infty}}/[32(kz)^3]}{\langle {S_{\alpha,\infty}} \rangle} \rightarrow \infty,~~\alpha=x,y
\label{eq:CRV_Sx}\\
\zeta_{S_z}(kz \gg 3/2) 
&= \frac{9 \, \sigma_{S_{z,\infty}}/[32(kz)^2]}{-\rmj \langle |\cE_0|^2 / \eta_0 \rangle / (kz)} 
= \rmj \frac{3 \sqrt{2}}{32 \, kz}
\label{eq:CRV_Sz}\\
\zeta_{S_t} (kz \gg 3/2) 
&= \frac{9 \, \sigma_{S_{t,\infty}}/[32(kz)^3]}{\langle {S_{t,\infty}} \rangle} \rightarrow \infty
\label{eq:CRV_St}\\
\zeta_{S}(kz \gg 3/2) &= \frac{3 \, \sigma_{S_\infty}/[16(kz)^2]}{-\rmj \langle | \cE_0|^2 / \eta_0 \rangle /(kz)} = \rmj \frac{\sqrt{6}}{16 \, kz}
.
\label{eq:CRV_S}
\end{align}
The decrease of $\zeta_{S_{(z)}}(kz)$ according to $(kz)^{-1}$ signifies that the nonlocal range of $\sigma_{S_z}$ decays more rapidly with $kz$ than that of $\langle S_z \rangle$ for $kz \gg 3/2$. At these large distances, it follows that $|\zeta_{S_{(z)}}(kz \gg 3/2)| \ll 1$.

Similarly, for the electric or magnetic energy densities
\begin{align}
\zeta_{U_\alpha}(kz \gg 3/2) &= \frac{3 \,  \sigma_{U_{\alpha,\infty}}/(2kz)}{3 \, \langle U_{\alpha,\infty} \rangle/(2kz)} = 1,~~\alpha=x,y
\label{eq:CRV_Ux}\\
\zeta_{U_z}(kz \gg 3/2) &= \frac{3 \, \sigma_{U_{z,\infty}}/[2(kz)^2]}{3 \, \langle U_{z,\infty} \rangle/[2(kz)^2]} = 1
\label{eq:CRV_Uz}\\
\zeta_{U_t}(kz \gg 3/2) &= \frac{3 \, \sigma_{U_{t,\infty}}/(2kz)}{3 \, \langle U_{t,\infty} \rangle/(2kz)} = \frac{1}{\sqrt{2}}
\label{eq:CRV_Ut}\\
\zeta_{U}(kz \gg 3/2) &= \frac{\sigma_{U_\infty}/(kz)}{\langle U_{\infty} \rangle/(kz)} = \frac{1}{\sqrt{3}}
\label{eq:CRV_U}
.
\end{align}
Unlike $\zeta_{S_{(\alpha)}}$, all $\zeta_{U_{(\alpha)}}$ are finite and independent of $kz$.
Note that $\zeta_{U_{(\alpha)}}(kz \gg 3/2) = \nu_{U_{(\alpha,)\infty}}$ even though all $\sigma_{U_{(\alpha)}}(kz)$ exhibit oscillations, as seen in Fig. \ref{fig:stdUe}.
All $\zeta_{U_{\rmem(,\alpha)}}(kz)$ are infinite because all $\langle U_{\rmem(,\alpha)}(kz) \rangle$ are independent of $kz$.

\section{Poynting's Theorem for RMS Fluctuations\label{sec:Poynting}}
Poynting's theorem of energy conservation links the local power flux for deterministic fields to the energy imbalance as
\begin{align}
\ul{\nabla} \cdot \ul{S}(\ul{r}) = - \rmj \omega [ U_{\rmm}(\ul{r}) - U_{\rme}(\ul{r}) ] \stackrel{\Delta}{=}  - \rmj \omega U_{\rmm - \rme}(\ul{r}).
\label{eq:Poynting_deterministic}
\end{align}
For deterministic or random plane waves ($\ul{\nabla}=-\rmj \ul{k}$) with fixed $\ul{k}$, (\ref{eq:Poynting_deterministic}) reduces to $S_k(\ul{r}) = \rmc U_{\rmm}(\ul{r}) - \rmc U_{\rme}(\ul{r})$, expressing the $\ul{k}$-directed Poynting component $S_k \equiv \ul{S} \cdot \ul{1}_k$ as the imbalance between the magnetic and electric scalar power densities.
For stochastic fields, it was shown in \cite{arnaLMAM} that (\ref{eq:Poynting_deterministic}) also holds for the mean power and energy densities , i.e., 
\begin{align}
\langle \ul{\nabla} \cdot \ul{S}(kz) \rangle = \ul{\nabla} \cdot \langle \ul{S}(kz) \rangle = - \rmj \omega \langle U_{\rmm-\rme} (kz) \rangle 
.
\label{eq:Poynting}
\end{align} 
Complex conjugation of (\ref{eq:Poynting_deterministic}) followed by side-by-side multiplication and ensemble averaging for random fields shows that the second-order moments are correspondingly related as 
\begin{align}
\langle |\ul{\nabla} \cdot \ul{S}(kz) |^2 \rangle = \omega^2 \langle | U_{\rmm-\rme} (kz)   |^2 \rangle.
\label{eq:Poynting_sq}
\end{align}
With $\sigma_{U_{\rmm-\rme}} = \sigma_{U_{\rmem}}$ and (\ref{eq:sigma_Uem_infty}), Poynting's theorem for RMS fluctuations at any $kz$ follows from (\ref{eq:Poynting}) and (\ref{eq:Poynting_sq}) as
\begin{align}
\sigma_{\ul{\nabla}\cdot\ul{S}} (kz) = \omega \sigma_{U_{\rmm-\rme}}(kz)  = \omega \sigma_{U_{\rmem,\infty}}
\sqrt{
1 + j^2_0 + \frac{3 j^2_1}{2} + \frac{j^2_2}{2}
}
.
\label{eq:sigma_Poynting}
\end{align}

The average energy imbalance for arbitrary $kz$ \cite[eq. (47)]{arnaLMAM}
\begin{align}
\langle U_{\rmm-\rme}(kz) \rangle 
= \frac{2\, \varphi_0 \langle |\cF_0|^2 \rangle }{ 3 } (j_0 - 2 j_2)
\label{eq:mu_Ume}
\end{align}
is plotted in \cite[Fig. 5]{arnaLMAM}. Near the boundary,
\begin{align}
\langle U_{\rmm-\rme} (kz \ll \sqrt{5}) \rangle &\simeq \frac{2 \, \varphi_0 \langle |\cF_0|^2 \rangle}{3} \left [ 1 - \frac{6(kz)^2}{5} \right ] \label{eq:mu_Ume_smallkz}
\end{align}
i.e., quadratically approaching $\langle U_{\rmm,x} (0)\rangle = \langle U_{\rme,z} (0)\rangle =\langle U_{\rmem,t} (0)\rangle =\langle U_{\rmem} (0)\rangle / 3$ for $kz \rightarrow 0$.
Its asymptotic envelopes are 
\begin{align}
{\mit \Xi}^\pm_{U_{\rmm-\rme}} (kz \gg 3/2) \simeq \pm \frac{\varphi_0 \langle |\cF_0|^2 \rangle }{ kz }
\label{eq:mu_Ume_largekz}
.
\end{align}

The CV of $U_{\rmm-\rme}$ (or $\rmc U_{\rmm-\rme}$) at arbitrary distances is 
\begin{align}
\nu_{U_{\rmm-\rme}}(kz) = \sqrt{\frac{3 \left (
1 + j^2_0 + \frac{3}{2}j^2_1 + \frac{1}{2} j^2_2 \right ) }{8 \left (j_0 - 2 j_2 \right )^2}}
.
\label{eq:nu_Ume_generalkz}
\end{align}
Its reciprocal is plotted in Fig. \ref{fig:nu}(a). From its range of values across $kz$, it can be inferred that, even though Poynting's law holds equally between the average power flux and the average energy imbalance as well as between their respective RMS fluctuation levels, the  exchanges between the fluctuations are stronger, i.e., $\sigma_{U_{\rmm-\rme}}(kz) \gg \langle U_{\rmm-\rme}(kz) \rangle$ for general $kz$. This is {\it a fortiori} the case for $kz \gg 1$, where $\langle \ul{S} \rangle$ and hence $\ul{\nabla} \cdot \langle \ul{S} \rangle$ approach null. Consequently, $\nu_{U_{\rmm-\rme}}(kz)$ approaches infinity for $kz \rightarrow +\infty$, also at quasi-periodic finite distances.

On the boundary, $\nu_{U_{\rmm-\rme}}$ reaches its minimum value $\sqrt{3}$ that exceeds $1$, i.e., the RMS fluctuation of the EM energy imbalance always remains larger than the mean imbalance, further increasing quadratically with distance as
\begin{align}
\nu_{U_{\rmm-\rme}} (kz \ll \sqrt{5}) &\simeq \sqrt{3} \left [ 1 + \frac{7 (kz)^2}{10}  \right ].
\label{eq:invnuUme}
\end{align}

From 
(\ref{eq:sigma_Uem_env}), (\ref{eq:sigma_Uem_env_min}) and (\ref{eq:mu_Ume_largekz}), the CRV $\zeta_{U_{\rmm-\rme}}(kz \gg 3/2)$ varies as $(kz)^{-1}$ and coincides with $| \zeta_{S}(kz \gg 3/2)|$, i.e.,
\begin{align}
\zeta_{U_{\rmm-\rme}}(kz \gg 3/2)
&= \frac{3 \, \sigma_{U_{\rmem,\infty}} / [8(kz)^2 ] }{ 2 \,\varphi_0 \langle | \cF_0|^2\rangle / (kz)} \nonumber\\
&= \frac{\sqrt{6}}{16 \, kz} 
= \left | \zeta_{S}(kz \gg 3/2) \right |.
\label{eq:zeta_Uem}
\end{align} 

\section{Conclusion}
In this article, the scalar EM power density $\rmc U_{\rmem}$ as a converted and scaled EM energy density and  the Poynting vector $\ul{S}$ for harmonic random fields were analyzed.
The divergence of $\ul{S}$ is proportional to $U_\rmm - U_\rme$, whereas $\rmc U_{\rmem}$ is  proportional to $U_\rmm + U_\rme$. 
For deterministic plane waves ($\ul{E} = \eta_0 \ul{H} \times \ul{1}_k$, $\ul{k} \cdot \ul{E} = 0$), both power quantities yield the same value.
In random non-plane wave environments, only the latter produces a nontrivial mean value in free space. 
The presence of an ideal PEC boundary affects both their means and standard deviations in different ways; cf. (\ref{eq:avgSz}) vs. (\ref{eq:avgUeUm}) and (\ref{eq:sigma_S}) vs. (\ref{eq:sigma_Uem}). For random fields near a PEC boundary, $U_{\rme}(kz) \not = U_\rmm(kz)$ qualitatively and quantitatively.
The results indicate the need for evaluating both $\ul{E}$ and $\ul{H}$ to obtain statistics of random EM power or energy density, unlike for deterministic plane waves where knowledge of either field suffices.

For Cartesian, tangential and full vectorial power and energy densities near a PEC boundary, their local standard deviations were obtained as 
eqs.
(\ref{eq:sigma_Sx}),
(\ref{eq:sigma_Sz}),
(\ref{eq:sigma_St}),
(\ref{eq:sigma_S});
eqs.
(\ref{eq:sigma_Ux}),
(\ref{eq:sigma_Uz}),
(\ref{eq:sigma_Ut}),
(\ref{eq:sigma_U}); and
eqs.
(\ref{eq:sigma_Uemx})--(\ref{eq:sigma_Uem}), respectively.
Similar to the averages, these standard deviations exhibit power-law type decay with the electrical distance $kz$, on average, at different rates. 
For the EM power, the envelope of $\sigma_{S_x}$ converges most rapidly to the free-space value, decaying as $(kz)^{-3}$.
For the energy densities $U_{\rme}$ and $U_{\rmm}$, the most rapid decay as $(kz)^{-2}$ occurs for the normal components $\sigma_{U_z}$; cf. (\ref{eq:sigma_Uz_env}). 
For 3-D vector fields, the standard deviation for power decays according to $(kz)^{-2}$ in (\ref{eq:asympenv_S_largekz}), which is more rapidly than $(kz)^{-1}$ for energy in (\ref{eq:asympenv_U_largekz}).  
Only the normal components $\sigma_{S_z}$ and $\sigma_{U_z}$ exhibit the same $(kz)^{-2}$ decay. 
Thus, the standard deviations exhibit an equal or sharper roll-off and narrower transition zone of $kz$ than the corresponding averages.
Near the boundary, $\sigma_{S_z}$ increases linearly with $kz$, while all other standard deviations vary quadratically, to leading order.

On approaching the PEC boundary, for $kz \ll \sqrt{5}$, the absolute RMS fluctuations of the normal and full scalar vs. Poynting powers diverge, i.e., $\sigma_{\rmc U_{\rmem(,z)}} > \sigma_{S_{(z)}}$ (Fig. \ref{fig:sigmaSxSzStot}). As a consequence, this may affect criteria for acceptable levels of field nonuniformity in MSRC validation procedures \cite{iec}, depending on whether dipole antennas or sensors vs. aperture antennas are used, when extending the working volume to include locations within the boundary zone. The relative RMS fluctuation of $S$ is much larger than for $\rmc U_{\rmem}$, at any $kz$; cf. (\ref{eq:nu_cUem_vs_S}). On the boundary, the CV of $S_z$ is finite, unlike the CV of $S$; cf. (\ref{eq:nu_Sz_smallkz}). 
For $kz \gg 3/2$, the decay of $\zeta_{S_z}$ occurs according to $(kz)^{-1}$ in (\ref{eq:CRV_Sz}), indicating that the range of nonlocal variation of $\sigma_{S_z}(kz)$ becomes progressively weaker with distance than that for $\langle {S_z}(kz)\rangle$. By contrast, the CRVs for energies in (\ref{eq:CRV_Ux})--(\ref{eq:CRV_U}) are constant and coincide with their respective CVs. 

The Poynting theorem, valid for averages of random power flux and random energy imbalance \cite{arnaLMAM}, extends to (\ref{eq:sigma_Poynting}) for their standard deviations. 
The RMS fluctuations of the power flux and imbalance are larger than their mean values, as expressed by (\ref{eq:nu_Ume_generalkz}) and} (\ref{eq:invnuUme}).


\begin{thebibliography}{99}
\bibitem{arnaLMAM} L. R. Arnaut and G. Gradoni, ``Average linear and angular momentum and power of random fields near a perfectly conducting boundary,'' \it IEEE Trans. Electromagn. Compat., \rm vol. 62, no. 4, pp. 1118--1127, Aug. 2020.

\bibitem{deleo2020} A. De Leo, G. Cerri, P. Russo, and V. Mariani Primiani, ``A novel emission test method for multiple monopole source stirred reverberation chambers,'' \it  IEEE Trans. Electromagn. Compat., \rm vol. 62, no. 5, pp. 2334--2337, Oct. 2020.

\bibitem{smith2012} S. M. Smith and C. Furse, ``Stochastic FDTD for analysis of statistical variation in electromagnetic fields,'' \it IEEE Trans. Ant. Propag., \rm vol. 60, no. 7, pp. 3343--3350, Jul. 2012.

\bibitem{hill1998} D. A. Hill, \it Electromagnetic Theory of Reverberation Chambers. \rm NIST Techn. Note 1506, U.S. Dep. Commerce, Boulder, CO, Dec. 1998.

\bibitem{dunn1990} J. M. Dunn, ``Local, high-frequency analysis of the field in a mode-stirred chamber,'' \it IEEE Trans. Electromagn. Compat., \rm vol. 32, no. 1, pp. 53--58, Feb. 1990.

\bibitem{hill2005} D. A. Hill, ``Boundary fields in reverberation chamberx,'' \it IEEE Trans. Electromagn. Compat., \rm vol. 47, no. 2, pp. 281--290, May 2005.

\bibitem{arnaPREmar2006} L. R. Arnaut, ``Spatial correlation functions of inhomogeneous random electromagnetic fields,'' \it Phys. Rev. E, \rm vol. 73, no. 3, 036604, Mar. 2006.

\bibitem{arnaRS2007} L. R. Arnaut, ``Probability distribution of random electromagnetic fields in the presence of a semi-infinite isotropic medium,'' \it Radio Sci., \rm vol. 42, RS3001, 2007.

\bibitem{iec} IEC Joint Task Force CISPR/A-SC77B: IEC 61000-4-21 Electromagnetic
Compatibility (EMC) -- Part 4-21: Testing and Measurement
Techniques -- Reverberation Chamber Test Methods, International
Electrotechnical Commission, 2nd ed., Jan. 2011.

\bibitem{arnaTEMCmay2006} L. R. Arnaut and P. D. West, ``Electromagnetic reverberation near a perfectly conducting boundary,'' \it IEEE Trans. Electromagn. Compat., \rm vol. 48, no. 2, pp. 359--371, May 2006.

\bibitem{arnaTQE2} L. R. Arnaut, ``Measurement uncertainty in reverberation chambers: I. Sample statistics,'' \it NPL Report TQE2, \rm 2nd ed., Nat. Phys. Lab., Teddington, U.K., Dec. 2008.

\bibitem{serr2020} R. Serra and C. Carobbi, ``A probabilistic interpretation of the IEC 61000-4-21 threshold levels for field uniformity in ideal reverberation chambers,''  \it Proc. 2020 EMC Europe Int. Symp. Electromagn. Compat., \rm 23--25 Sep. 2020, Rome, Italy.

\bibitem{yagl1962} A. M. Yaglom, \it An Introduction to the Theory of Stationary Random Functions. \rm Prentice--Hall: Englewood Cliffs, N. J., 1962.

\bibitem{king2009} F. W. King, \it Hilbert Transforms. \rm Cambridge University Press: Cambridge, U.K., 2009.

\bibitem{arnaCE} L. R. Arnaut, ``Compound exponential distributions for undermoded reverberation chambers,'' \it IEEE Trans. Electromagn. Compat., \rm vol. 44, no. 3, pp. 442--457,  Aug. 2002.
\end{thebibliography}
\end{document}